\documentclass[useAMS,usenatbib]{mn2e}
\usepackage{graphicx,hyperref, subfigure}
\usepackage{xcolor}

\newcommand\aj{{AJ}}%
\newcommand\araa{{ARA\&A}}%
\newcommand\apj{{ApJ}}%
\newcommand\apjl{{ApJ}}%
\newcommand\apjs{{ApJS}}%
\newcommand\aap{{A\&A}}%
\newcommand\aaps{{A\&AS}}%
\newcommand\mnras{{MNRAS}}%
\title[Filaments]{Filament Identification Through Mathematical Morphology}
\author[Koch \& Rosolowsky]{Eric W. Koch$^{1,2}$\thanks{E-mail:
ekoch@ualberta.ca (EWK); erosolow@ualbeta.ca (EWR)} and Erik W. Rosolowsky$^{1,2}$\footnotemark[1]\\
$^{1}$University of Alberta, Department of Physics, 4-183 CCIS, Edmonton AB T6G 2E1, Canada\\
$^{2}$University of British Columbia, Okanagan
  Campus, Department of Physics, 3333 University
  Way, Kelowna BC V1V 1V7 Canada}
\begin{document}

\date{Draft date: \today}

\pagerange{\pageref{firstpage}--\pageref{lastpage}} \pubyear{2015}

\maketitle

\label{firstpage}

\begin{abstract}
We present a new algorithm for detecting filamentary structure {\it FilFinder}. The algorithm uses the techniques of mathematical morphology for filament identification, presenting a complementary approach to current algorithms which use matched filtering or critical manifolds.  Unlike other methods, {\it FilFinder} identifies filaments over a wide dynamic range in brightness.  We apply the new algorithm to far infrared imaging data of dust emission released by the {\it Herschel} Gould Belt Survey team.  Our preliminary analysis characterizes both filaments and fainter striations.  We find a typical filament width of 0.09 pc across the sample, but the brightness varies from cloud to cloud.  Several regions show a bimodal filament brightness distribution, with the bright mode (filaments) being an order of magnitude brighter than the faint mode (striations).  Using the Rolling Hough Transform, we characterize the orientations of the striations in the data, finding preferred directions that agree with magnetic field direction where data are available.  There is a suggestive but noisy correlation between typical filament brightness and literature values of the star formation rates for clouds in the Gould Belt.
\end{abstract}


\section{Introduction}

The structure of molecular clouds (MCs) is thought to guide the star formation process, establishing both the rate and location of star formation.  Unfortunately, the molecular hydrogen that makes up the majority of the mass in  MCs has negligible emission at the typical 10~K temperatures that characterize the bulk of MCs \citep{psp3}.  Furthermore, high dust extinction precludes H$_2$ absorption line mapping of cloud structure over large areas \citep{spitzer73}.  Thus, studies of MC structure must rely on other tracers.  Historically, the most successful of these methods have been molecular line tracers, notably the rotational lines of CO and its isotopologues, infrared dust emission, and dust extinction in the near infrared and optical.

Much of the early work studying cloud structure emphasized data from large area mapping of molecular emission lines \citep[e.g.,][]{bally-orion}.  Since the lines are velocity-resolved, the additional velocity dimension provided essential context for understanding the dynamics of MCs.  Unfortunately, the molecular line tracers are biased by both chemistry, excitation and radiative transfer effects.  These effects combine to limit the dynamic range of a given molecular line tracer to a relatively narrow range in density \citep{evans-araa}.    The past decade has seen a shift in attention toward structure traced by dust, enabled by the thermal emission maps provided by the {\em Spitzer} and {\em Herschel} missions.  Dust emission provides what appears to be a less biased tracer of cloud structure, though variations in dust temperature, abundance and emissivity still require careful treatment \citep[e.g.,][]{schnee10}.  Complementary work has been completed using dust extinction in the near infrared, though these efforts have largely focused on nearby clouds \citep[e.g.,][]{nice-method,nicer}.

Qualitative analyses of these tracers forwarded a rich view of the structure in MCs, including the significance of filamentary structure \citep[e.g.,][]{lynds62, schneider79, bally-orion, falgarone-irammaps, nagahama-orion}.  However, a quantitative analysis of these data was governed by the statistical and numerical techniques adopted and these approaches shaped the interpretation of MC structure.  These analyses usually followed one of two approaches: statistical measurements of cloud image properties or the generation of object catalogs.   The statistical image characterization includes methods like Fourier or wavelet analysis which are usually anchored in the theory of turbulence \citep{stutzki98,wavelet-clouds}.  These methods are most useful when the measurement can be linked directly back to the underlying physical properties of the gas.  Statistical methods have consistently emphasized the turbulent nature of MCs and recent work has emphasized retrieving the underlying properties of turbulence from the observations \citep[e.g.,][]{vca,lp04}.  In contrast, the object cataloging approach splits the structure of MCs into an ensemble of substructures and explore the properties of that ensemble.  The classic two algorithms in the field are Gaussclumps \citep{gaussclumps2} and Clumpfind \citep{clumpfind}.  These methods are designed around exploring the structure seen in molecular spectral line data cubes. The algorithms search for objects that have low (three-dimensional) aspect ratios, either by construction (Gaussclumps) or implicitly given the underlying pixel grid (clumpfind).  Similar methods developed for two-dimensional data have been applied to continuum maps of the ISM \citep{sextractor,bgps-paper2}.  The view of the molecular ISM derived from these object identification approaches is frequently described as {\em clumpy}. This clumpy view is drawn from the analysis of wide area mapping of MCs restricted to using only CO as a tracer and mapped at relatively poor quality compared to the current state of the art: for example, compare data in \citet{clumpfind} to \citet{taurus-fcrao}. This view has been emphasized by algorithmic construction and analysis. These algorithms adapt poorly to recovering structures with large aspect ratios, particularly when the structure is not straight in the plane of the sky.  Additionally, using CO isotopologues as molecular gas tracers limits the dynamic range to roughly an order of magnitude in volume density for each tracer \citep{evans-araa}.  This density sensitivity obscures filamentary structure with a wide range of densities along a line of sight.

While large-area molecular line mapping now provides data sets with wide dynamical range, the observations are relatively time consuming.  Compared to CO observations, dust emission mapping is fast and relatively high resolution. The recent dust emission data, most notably the {\em Herschel} mapping of dust emission in the far infrared and submillimetre (70 to 500 $\mu$m) present wide area maps with good dynamic range and complementary observational biases with respect to molecular line mapping.  However, dust maps integrate over the line of sight, potentially mixing regions with different physical conditions.  With the kinematic information from CO line emission, some of these overlaps can be resolved though line-of-sight projections still influence the interpretation \citep{beaumont13}.

Initial analysis of dust emission maps has focused on the {\em filamentary} structure of the MC, specifically noting features in the images with large aspect ratios \citep{gbs-andre,pp6-andre}.  Several qualitative descriptions of the filamentary networks have been proposed including a pervasive hub-filament structure \citep{myers-filament},  converging filamentary patterns \citep{peretto-pipe}, or parallel networks \citep{busquet13}.  Filaments appear to be ubiquitous though their qualitative interpretation can be subject to human judgement and the tracers used.  While the dust emission data highlight the two-dimensional column density distributions, kinematic information from dense gas tracers can be used to trace the gas motion.  Several individual filaments and hub-filament structures have been analyzed in detail and have even been argued to be the fundamental structures in the cluster formation process \citep[e.g.,][]{myers-filament,hkirk-serpens}.

While providing essential guidance, these single-object studies must be paired with broader population studies to place their results in context.  We require a robust characterization of filamentary structure that can be used to quantify the properties of filaments for an ensemble of clouds across the ISM.  With such a quantification of the ensemble, the population can be checked for variation with physical parameters and compared to both individual cases and simulations of cloud structure.  This need for a systematic characterization of filament properties has driven the development or adaptation of several new cataloging algorithms in the study of ISM structure.  These algorithms approach the problem from different perspectives.

The DisPerSE algorithm \citep{disperse} was originally developed for identifying structures in three-dimensional cosmological simulations.  DisPerSE leverages computational Morse theory to identify the critical surfaces and subsequently draw smooth one-dimensional manifolds (filaments) through a density field.  At the heart of the method lies a Delaunay triangulation of the domain which serves as a well defined set of points where the gradients of the underlying field can be rigorously evaluated.  The domain is partitioned into regions defined by these critical points and the filamentary structures are identified at the borders of these domains.  By comparing the robustness of the definition subject to the variation of the points that define the field, the algorithm is made robust in the presence of sampling noise.  The algorithm works in arbitrary dimension and can be applied to both two-dimensional dust emission data and three-dimensional spectral line data cubes.

Following a similar approach, the Hi-GAL survey team \citep{higal-survey} is exploring the use of image gradients and the local Hessian matrix as a basis for identifying filamentary structure \citep{schisano-2014}.  The eigenvalues of the Hessian matrix identify the direction of elongation and can find those sets of points on ridges for which the curvature is large and negative in one direction and small in the other.  These sets of points define regions which are aggregated into filaments for which the properties can be characterized. The Hi-GAL approach then processes the Hessian map using image processing approaches similar to those proposed in Section \ref{section:methods} below. Since the Hessian matrix can be generalized to arbitrary dimension, this approach may be generalizable to spectral line data, but development to date is focused on the two-dimensional images.

\citet{salji15} also present a Hessian-based approach but instead uses a multiscale analysis to enhance filamentary structures with a range of widths relative to the filter size.

Finally, the {\it getfilaments} algorithm \citep{getfilaments} identifies filaments through their persistence in the image data across multiple spatial scales.  The algorithm progressively smooths and subtracts off signal at different spatial scales.  The smoothing kernel is a Gaussian and increases in size exponentially up to the maximum scale probed.  The decomposition is similar to the continuous form of the {\em a trous} wavelet algorithm \citep{starck-murtagh}.  Filaments are identified as significant structures with lengths minimally changed by the convolutions but widths equal to the convolution beam.  The algorithm is then applied to the data across several wavebands, and filaments are reconstructed based on a common appearance in several images.

In this paper, we present a separate approach to identifying filaments based on the application of mathematical morphology \citep{mathematical-morphology}.  Mathematical morphology has developed a suite of operations for digital image processing based on the underlying pixel grid of the image, which is usually rectilinear, and the discrete values of the image data.  In particular, these operators act on grey-valued images, with integer value from 0 to $2^k-1$ where $k$ is usually 1 (binary masks), 8, 16 or 32.  We identify regions of emission and find their skeletons to identify filamentary structures.  We then prune these skeletons to produce a robust catalog of filaments.

We present our algorithm below.  In Section \ref{section:methods} we outline the algorithm and apply the algorithm to the {\em Herschel} Gould Belt Survey (Section \ref{section:data}) to generate a catalog of filamentary structures (Section \ref{section:results}).  We discuss our results in Section \ref{section:discussion}.

\section{Methods}
\label{section:methods}

Identifying and segmenting filamentary structure is a challenging computational problem because there is little prior information about the structure of the features in an image. This section describes our method of segmenting filamentary structure using intensity data and how the main properties of each structure are calculated.

This new algorithm builds off a rich suite of existing methods developed in the field of {\it mathematical morphology}, which is used across science for image processing.  Our treatment relies on a well developed theoretical framework that is laid out in, for example, \citet[][ hereafter S09]{mathematical-morphology}.  Such methods are well validated and public domain implementations are readily available\footnote{For example, we rely on \url{http://scikit-image.org/}}.  We find that this method is robust to changes in algorithmic parameters and the approach can identify filamentary structure over several orders of magnitude in intensity.  We refer to our algorithm as {\it FilFinder}; it is implemented in Python and is publicly available\footnote{\url{https://github.com/e-koch/FilFinder}}.

\subsection{Isolating Filamentary Structure} 
\vspace{0.01in}
\label{sub:isolating_filamentary_structure}
\begin{figure*}
\includegraphics[width=0.9\textwidth]{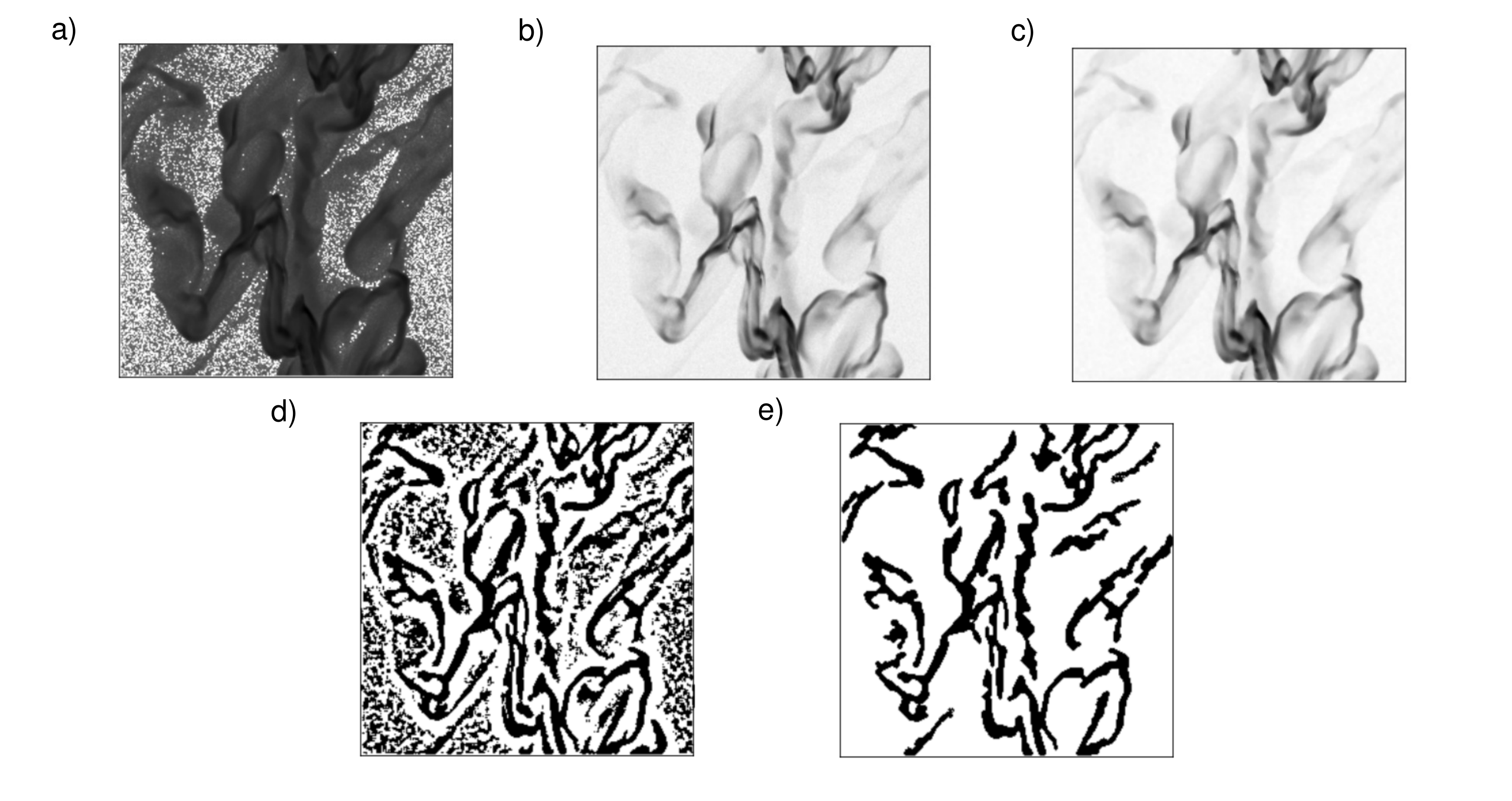}
\caption{\label{fig:steps} Steps in the masking process. a) Simulated data with Gaussian noise shown on a logarithmic intensity scale. b) Flattened using an arctan transform. c) Smoothed using a Gaussian filter. d) Adaptive thresholding. e) Remove unphysical, small objects.}
\end{figure*}

The algorithm begins with a five-step preprocessing sequence that identifies regions in the image containing filamentary structure.  These steps are demonstrated in Figure \ref{fig:steps}, which shows the application of these steps to simulated data.  The image is drawn from an Enzo simulation \citep{enzo} of the column density in a stirred turbulent box with initial conditions chosen to mimic interstellar clouds and produce filamentary structure.  We use simulated data to demonstrate the algorithm since the results can be produced free from observational noise and then subsequently degraded to establish the influence of observational effects.  First, we fit a log-normal distribution to the brightness data, identifying the mean $\mu$ and standard deviation $\sigma$ of the log-intensity. The image is flattened using an arctangent transform: $I' = I_0 \mathrm{arctan}(I/I_0)$ where the normalization $I_0\equiv\exp(\mu + 2\sigma)$.  Since $\mathrm{arctan}(x)\sim x$ for $x < 1$, the effect of the transform is to leave emission below the threshold minimally changed but values significantly above the threshold asymptotically approach $\pi I_0/2$.   This transform suppresses objects that are significantly brighter than the filamentary structure within the image (usually bright cores). We note that the final mask does not change significantly for small changes in $I_0$. We discuss this effect in Appendix \ref{app:parameter_sensitivity}. Second, the flattened data are  smoothed with a Gaussian that has a FHWM of 0.05 pc -- half the expected width of a filaments found by \citet{arzou2011_ic5146}. The smoothing minimizes variations within the filaments, which eliminates spurious elements later in the process, while maintaining the large-scale structure. Third, we create a mask of the filamentary structure by applying an adaptive threshold to the smoothed image. This is the crucial step in creating the filament mask. Adaptive thresholding considers each pixel in the image in comparison to a patch of the image around each pixel and uses a criterion to keep or discard that pixel. The criterion used here is that the intensity of the central pixel must be greater than the median of the neighborhood around it. The local threshold is what allows for filamentary structure to be detected over the entire dynamic range in brightness within the the image. We choose a patch size of 0.2 pc, as twice the expected filament width \citep{arzou2011_ic5146}. We note that the choice of patch size affects the filamentary mask shape, but this patch size is not used when calculating the filament widths (see Section \ref{sub:width}). We combine this mask with a globally thresholded mask, which removes regions below the noise level in the image.

These first three steps are sensitive to the chosen normalization value for the arctan transform ($I_0$) and the patch size used for the adaptive threshold. The vast majority of the structure is recognized for a range of reasonable values of both parameters, however, the connectivity between detected regions can change drastically. We find that optimal results are obtained by setting these parameters to the values explained above but we explore the effects of these choices in Appendix \ref{app:parameter_sensitivity}.  Since mathematical morphology methods are based on the geometry of the image pixels, some instability can arise when the physical scales that we use to estimate the algorithm parameters map to small pixel sizes.  In particular, we find that for regions where the adaptive threshold patch size is smaller than 40 pixels across, spurious fragmentation occurs.  In these cases, the algorithm interpolates the image onto a larger grid for the first 3 steps. The returned mask is then regridded back onto the original image size.

The remaining two steps clean small, spurious features from the adaptive threshold mask. Objects are removed that have an area smaller than the typical smallest filamentary object: $5\pi(0.1 \textrm{pc})^2$, where 0.1 pc is a typical filament width and 5 is the aspect ratio for an object to be considered a filament. This definition was chosen empirically after examining the output of multiple datasets. The vast majority of filaments make up a network, and so they deviate significantly from the elliptical description of filaments. We found that, for the dataset analyzed in this work, regions with a smaller area than the threshold led to failed fits to their radial profiles (Section \ref{sub:width}), consistent with these being spurious features.  Finally, a median filter with a size of 0.05 pc is applied to the mask, which removes small features from the edges of objects.  This step limits spurious branches in the filament identification steps that follow.


\subsection{Tracing Filaments using Skeletons} 
\label{sub:analyzing_filament_properties}

The combination of these five pre-processing steps establish the mask used for subsequent analysis.   Each structure within the mask is reduced to a skeleton using a Medial Axis Transform \citep{blum67}. This method reduces an object to skeleton form (i.e., one pixel wide) such that it is minimally connected. The Medial Axis Transform calculates the position of the skeleton by computing the minimum distance of each pixel in the object to a pixel outside of it. A skeleton pixel is one which maximizes this distance with respect to its neighbouring pixels along one direction (e.g., vertical), but does not in the other direction (e.g., horizontal).  This can be thought of as finding the circle with the maximum radius that will fit into the object. As a simplified example, the skeleton of a circular object would simply be the pixel at the centre. If the object were then stretched in the vertical direction, its skeleton would be a vertical line. Thus the skeleton pixels are points that are the centres of  the maximal circles that fit into the object.

\begin{figure}
\includegraphics[width=0.48\textwidth]{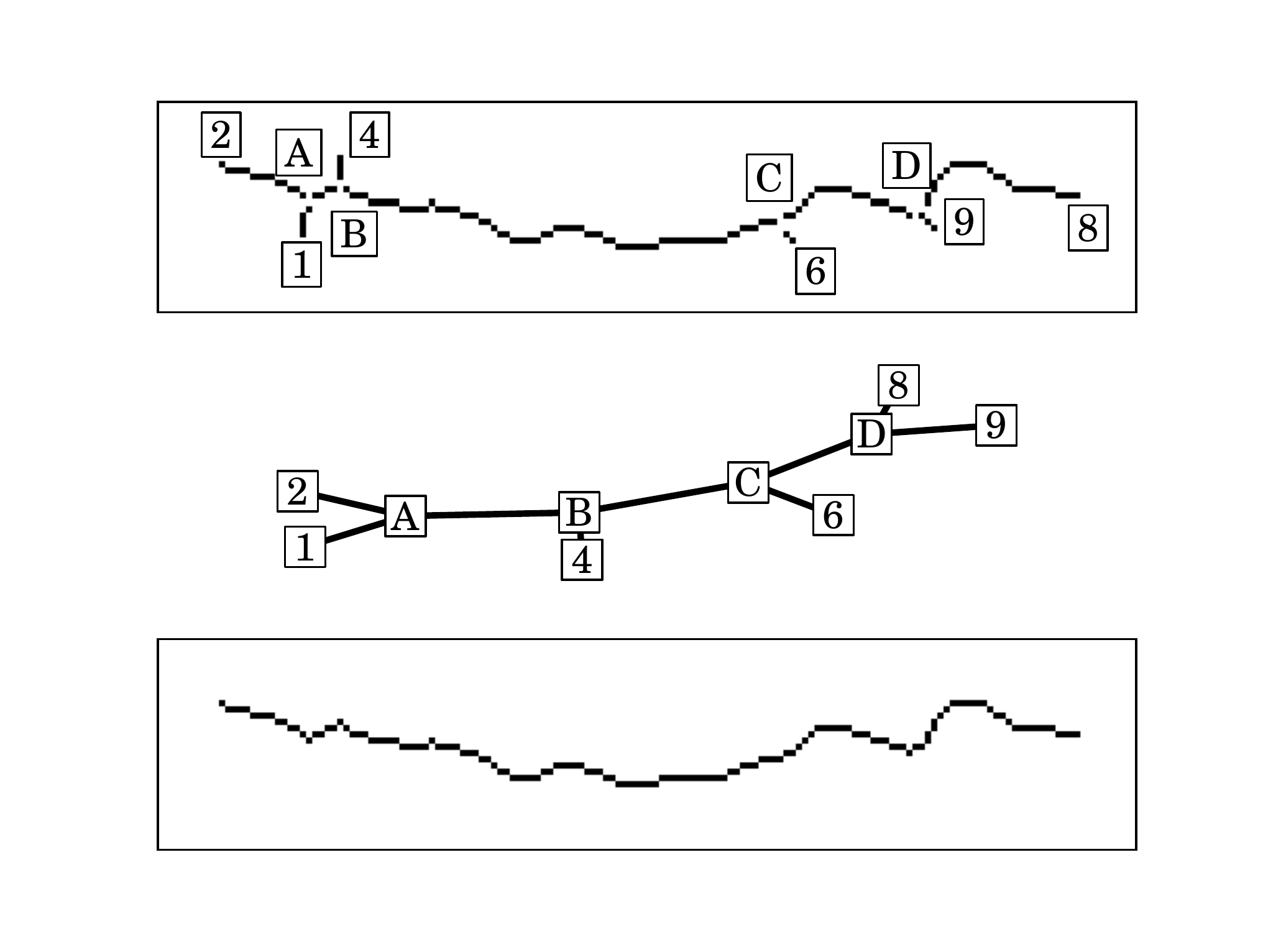}
\caption{\label{fig:skel} Top: An un-pruned skeleton with intersection points removed. Numerical labels denote branches and alphabetical ones represent intersections. Middle: The skeleton converted to a graph. Bottom: The final skeleton after the pruning process.}
\end{figure}

After the Medial Axis Transform, we process the resulting skeleton into the filamentary structures.  The main steps in this process are illustrated in Figure \ref{fig:skel}.  Initially, the skeleton contains many small branches resulting from small deviations in the mask boundary that get amplified under the transform (Figure \ref{fig:skel}, top panel).  These typical features need to be ``pruned'' leaving the dominant skeleton feature behind for characterization.  Unfortunately, most pruning methods present in the literature (S09) shorten the length of the skeleton, which we regard as a scientifically relevant property of the filaments.   Instead, we complete a full analysis of the unpruned skeleton and use the length information in the skeleton to identify branches for pruning.   We first label each pixel in the skeleton by the number of adjacent neighbours.   There are three relevant cases: end points (1 neighbour), body points (2 neighbours) or intersections ($\ge 3$ neighbours).   Intersection points are removed from the skeleton, returning a collection of branches.

We exploit the fact that each branch must be comprised only of end and body points to calculate the length of a branch.  Each pixel is connected to at most two neighbours (which do not touch each other) and the length of each branch is unambiguously defined.  The skeleton is then converted to a weighted graph, where the nodes of the graph consist of the end and intersection points, and the edges are the branches.  Edges are weighted using a sum of weights derived from the branch length and the average intensity of the image along the branch. Both quantities are normalized by the sum over all branches in the skeleton, ensuring a weight between 0 and 1. A shortest-path algorithm is run on the graph, which identifies the shortest possible path between each pair of end points. From this set of paths, we identify the longest path through the skeleton. The sum of branch lengths for this longest shortest-path is defined to be the length of the filament.

Any remaining branches which are not part of the longest path are now subject to pruning.  A branch is pruned provided its deletion does not affect the skeleton's connectivity, its length is below a given threshold, and its average intensity is insignificant relative to the rest of the skeleton.  The length threshold is set to 3 times the beamwidth of the image; using a shorter threshold did not effectively prune spurious branches. The intensity threshold is set to 0.1 of the total intensity along the entire skeleton. The value of the intensity threshold was found to be insensitive to moderate changes, however it ensures the removal of long spurious features over background regions.

We note that {\it FilFinder} does not prohibit loops in the skeleton structure. While this can be an artifact of the medial axis transform, we find that the majority of the loops describe the actual pattern of emission. The shortest-path algorithm used to derive the filament length is not able to handle loops in the graph. To overcome this, we only include the path through the loop that has a higher weight when computing the length. The other portion of the loop, provided it does not alter the overall connectivity, is then eligible to be pruned. To eliminate loops from spurious holes in the mask, we fill in holes that have areas smaller than the beam area.


\subsection{Plane Orientation and Curvature} 
\label{sub:plane_orientation_and_curvature}

A novel method for parameterizing the orientation of linear structure known as the Rolling Hough Transform (RHT) was introduced by \citet{rollinhough}. We provide a basic explanation of the algorithm here. The RHT is based on the family of Hough transforms, which map $(x,y)$ space into ($r$, $\theta$) space using the equation for a straight line,
\begin{equation}
  \label{eq:linear}
  r = x\cos{\theta} + y\sin{\theta},
\end{equation}
\noindent
where $r$ is the radius from the origin and $\theta$ is the angle from a defined direction \citep{Duda:1972:UHT:361237.361242}. The RHT differs from this original Hough transform definition by restricting to $r=0$ and defining the origin to be at the centre of a disc of a given diameter. Then, $\theta=0$ is defined to be in the positive $y$-direction. The transform is performed by measuring the response within the disc while varying $\theta$ between 0 and $\pi$ at each specified pixel. The value of the transform around each pixel is the angle at which the transform has the largest response.

We perform the RHT using a disc with a diameter of 3 times the beamwidth at each pixel within a skeleton. The diameter was determined to be the smallest diameter which eliminated significant pixelization effects.  Smaller discs can become dominated by the axes and diagonal directions ($\theta= 0^{\circ}, 45^{\circ}, 90^{\circ}$) instead of the larger structure.  Discs significantly larger than filaments can wash out the directional information.  We sum the response of the transform from all pixels to arrive at a distribution of angles in the plane for the skeleton.  Unlike in \citet{rollinhough} where the expectation value of the RHT is compared to the starlight polarization angle, we do not have a quantity to set the zero of the distribution. Instead, we calculate the weighted directional mean and define it as the orientation of the filament,
\begin{equation}
  \label{eq:direc_mean}
  \langle\theta \rangle = \frac{1}{2} \tan^{-1}\left(\frac{\Sigma_i w_i\sin2\theta_i}{\Sigma_i w_i\cos2\theta_i}\right),
\end{equation}
\noindent
where $w_i$ is the normalized value of the transform at $\theta_i$ such that $\Sigma_iw_i=1$. Here $\theta$ is defined on $\left[-\pi/2, \pi/2\right)$. We calculate the confidence intervals on the directional mean using Equation 3 from \citet{fl-1983-direcstats},
\begin{equation}
  \label{eq:direc_ci}
  \langle\theta \rangle \pm \sin^{-1} \left( u_{\alpha} \sqrt{ \frac{1-\alpha}{2R^2} } \right)
\end{equation}
\noindent
where $u_{\alpha}$ is the $z$-score of the two-tail probability. The quantity $\alpha=\Sigma_i\cos{\left[2w_i\left(\theta_i-\langle\theta\rangle\right)\right]}$ is the estimated weighted second trigonometric moment and $R^2=\left[\left(\Sigma_iw_i\sin{\theta_i}\right)^2 +\left(\Sigma_iw_i\cos{\theta_i}\right)^2\right]$ is the weighted length of the vector.

\begin{figure}
\includegraphics[width=0.48\textwidth]{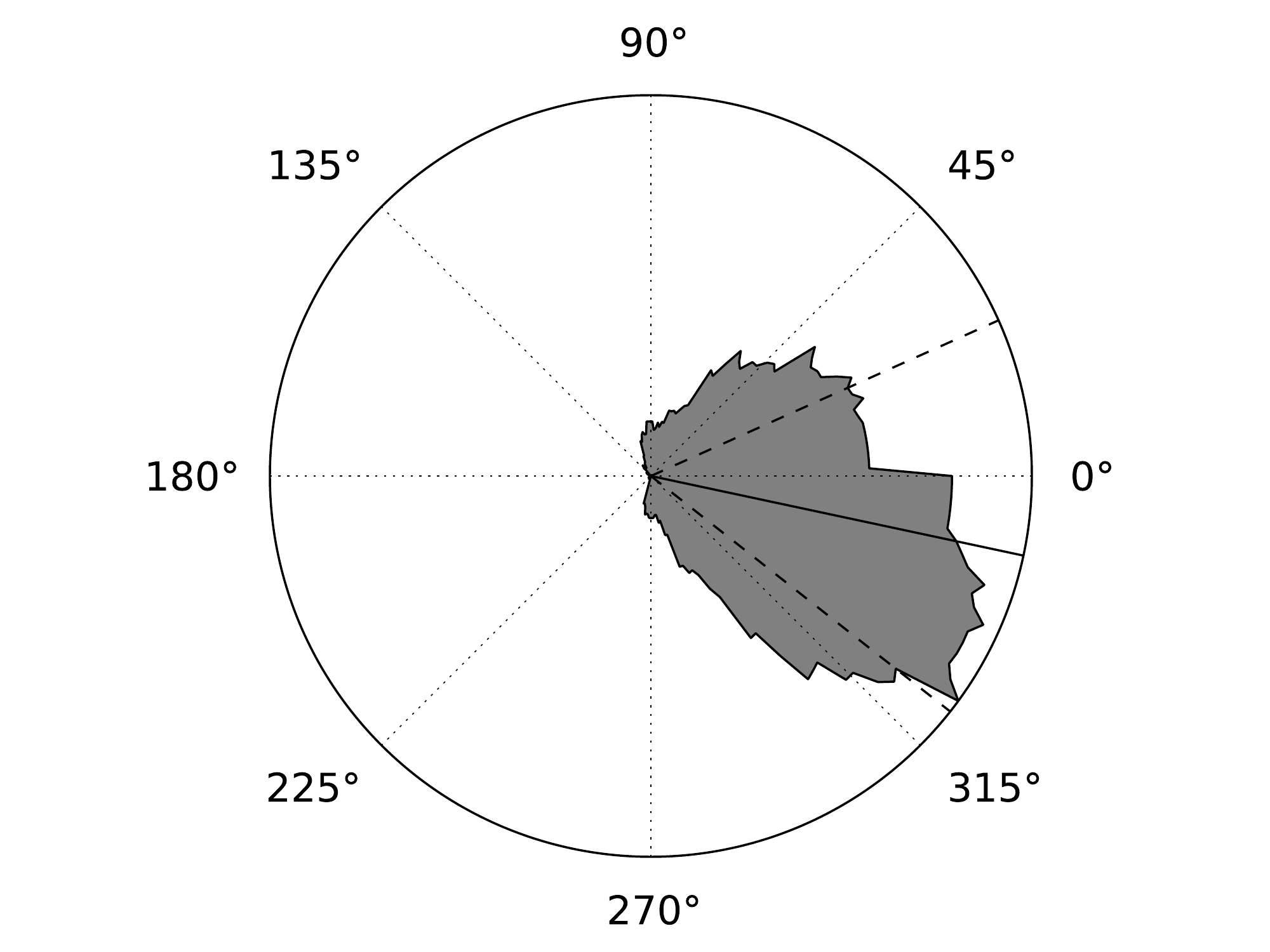}
\caption{\label{fig:rht} The output of the RHT for the skeleton shown in Figure \ref{fig:skel}. Note $2\theta$ is plotted to visualize the continuous boundary. The median is shown by the solid line and the quartiles by the dashed lines.  The curvature ($\delta \theta$) of the filament is defined the interquartile range, or the half the angle between the two dashed lines, where the `half' is introduced because we are plotting $2\theta$.}
\end{figure}

These quantities make the assumption that the distribution is unimodal, which does not hold for all filaments. Because of this, and since the confidence intervals are limited to the range of $\theta$, we define the curvature $\delta \theta$ conservatively to be the interquartile range (IQR, $z=0.67$). Filaments with a multi-modal response have an ill-defined orientation, however their dispersion is accounted for by the measure of curvature. We show an example of the distribution, angle of orientation, and curvature in Figure \ref{fig:rht}. Like the polarization angle of starlight, the orientation of a filament is ambiguous, so $\theta$ is confined to a range of $\pi$ radians.

\subsection{Width} 
\label{sub:width}

We calculate a filament's width by building a radial profile with respect to the skeleton. We consider a distance transform using the skeleton as the reference point.  To avoid counting pixels multiple times, we calculate a global distance transform for an entire image using all of the skeletons. A given pixel is then included in profile calculations only for the skeleton to which it is closest. The profiles are limited to a maximum distance of 0.3 pc to ensure profiles only account for the profile of the filament. We found that this threshold did not cut off any features of the profile in question for all regions that we analyzed (Section \ref{section:results}). We found that this limit was not adequate in many cases, particularly when filaments are closely packed together. To ensure we only fit the profile of the filament in question, radial profiles are further cut by removing portions where the profile begins to increase.

\begin{figure}
\includegraphics[width=0.48\textwidth]{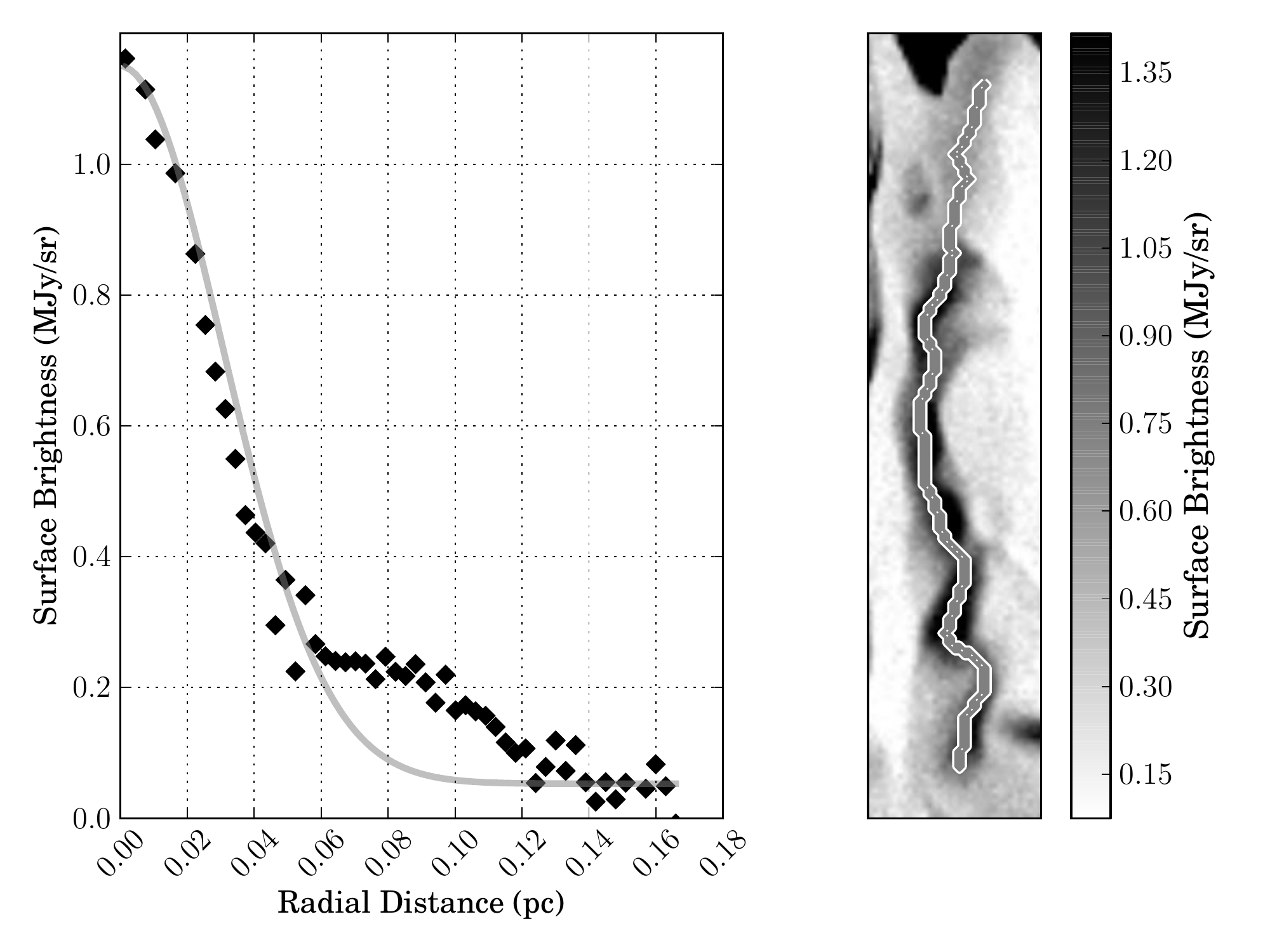}
\caption{\label{fig:width} Left: The radial profile of the filament shown to the right and in Figure \ref{fig:skel}. The profile is the result of averaging the intensity by binning pixels with respect to their distance from the skeleton. Right: Skeleton of the filament plotted over the image.}
\end{figure}

To derive the width and the surface brightness of that filament, we fit a Gaussian with a constant offset (background) to the radial profiles. The radial profile and Gaussian fit for a filament are shown in Figure \ref{fig:width}. We define the filament width to be the full-width-half-maximum (FWHM) value deconvolved by the beamwidth, as described by \citet{arzou2011_ic5146}.

Not all filaments showed a Gaussian profile, so to obtain some measure of filament width in these cases, we adopt a parallel, non-parametric method. We estimate the peak and background intensities in the profile using the 5th and 99th percentiles, respectively. We then interpolate the profile and calculate the distance where the intensity drops to half of the peak intensity relative to the background.  This value reproduces the Gaussian FWHM value for a Gaussian profile but still provides a representative width measurement for non-Gaussian filaments. The error is estimated by calculating the range of distance between the 45\% and 55\% percentiles in the profile.

The non-parametric method is used only in the cases where a Gaussian does not return a good fit. We quantify a `good fit' as one that has errors that do not exceed the absolute value of the parameters and the reduced $\chi^2$ value is below $\sim 10$. The exact cutoff value used is largely insensitive to determining the quality of near Gaussian profiles. Those which have significant, non-random deviations from a Gaussian return a significantly higher reduced $\chi^2$. This method is used as the last resort in an attempt to derive some useful information from the radial profile. In the results presented (Section \ref{section:results}), only 2$\%$ of the derived widths use the non-parametric approach.  The non-parametric method is used primarily in two situations: there is excessive crowding with other filaments leading to a lack of valid pixels for the profile, or a bright, compact feature not associated with the filamentary structure caused the profile to increase at greater distance, leading to a severely truncated profile. This latter effect occurs for short filaments, as the smaller number of surrounding pixels has a greater effect when averaging within each bin.

\section{Data}
\label{section:data}

\begin{table*}
 \label{tab:properties}
 \centering
 \begin{minipage}{100mm}
  \caption{Adopted Cloud Properties. The masses are calculated using the assumptions from Section \ref{sub:stability} and the Planck offsets.}
  \begin{tabular}{@{}lccccc@{}}
  \hline
   Name     &  Reference & Distance & $\Sigma_{\mathrm{SFR}}$ & $M_{\mathrm{cloud}}$ & Planck Offset\\
            &            & (pc)     & ($M_{\odot}$~yr$^{-1}$~pc$^{-2}$) & $(M_{\odot})$ &
            (MJy~sr$^{-1}$)\\
\hline
Aquila            & 1, 3    & 260 & 1.80 & 30\,000  & 85.5   \\
California Centre & 2, 3   & 450 & 0.32 & 23\,000 & 9.0    \\
California East   & 2, 3   & 450 & 0.32 & 12\,000 & 10.1   \\
California West   & 2, 3   & 450 & 0.32 & 5200 & 14.7   \\
Chamaeleon        & 4, 5, 3   & 170 & 3.88 & 1500 & -879.1 \\
IC-5146           & 6, 3   & 460 & 0.38 & 6800  & 20.7   \\
Lupus I           & 7, 3   & 150 & 0.37 & 900 & 14.4   \\
Orion-A Centre    & 8, 9, 10   & 400 & 0.48 & 12\,000  & 32.6   \\
Orion-A South     & 8, 9, 10   & 400 & 0.48 & 18\,000 & 35.2   \\
Orion-B           & 11, 10   & 400 & 0.084 & 44\,000& 26.2   \\
Perseus           & 12, 3  & 235 & 1.32 & 5100 & 23.7   \\
Pipe              & 13, 11  & 145 & 0.032 & 1300 & 31.7   \\
Polaris           & 1, 14  & 150 & 0.0 & 1000 & 9.3    \\
Taurus            & 15, 16, 17 & 140 & 0.147 & 2200 & 21.2   \\
\hline
\end{tabular}
References -- 1: \citet{gbs-andre}, 2: \citet{california_cloud}, 3: \citet{evans-2014-sfr}, 4: \citet{winston-2012-cha}, 5: \citet{chamaeleon-striations}, 6: \citet{arzou2011_ic5146}, 7: \citet{rygl-2013-lup}, 8: \citet{polychroni_orionA}, 9: \citet{roy-2013-oriona}, 10: \citet{lada-2010}, 11: \citet{schneider-2013-orionb}, 12: \citet{pezzuto-2012}, 13: \citet{pipe-filaments}, 14: \citet{miville-2010}, 15: \citet{kirk_taurus} 16: \citet{Palmeirim-2013-taur}, 17: \citet{heiderman-2010}.
\end{minipage}
\end{table*}

We apply our new filament extraction method to the observational data acquired as part of the {\it Herschel} Gould Belt Survey \citep[HGBS;][]{gbs-andre}.  This program surveyed 16 nearby molecular clouds using the PACS and SPIRE instruments \citep{spire-instrument}.  The survey produced maps of emission with wavelengths of 70 $\mu$m, 160$\mu$m, 250 $\mu$m, 350 $\mu$m and 500 $\mu$m, which primarily traces dust emission in these nearby clouds.  The {\it Herschel} data have provided a new view into the (dust) column density structure in nearby clouds and were arguably the main reason for the community's increasing interest in filamentary structure \citep{gbs-andre}.   We used the preliminary release data available from the HGBS survey webpage \footnote{\url{http://www.herschel.fr/cea/gouldbelt/en/Phocea/Vie_des_labos/Ast/ast_visu.php?id_ast=66}}.  The individual data have been described in a series of publications as given in Table \ref{tab:properties}, which also presents adopted cloud properties.  At present, the HGBS team has not released a final reference set of images, and our preliminary results described below should be revised once a reference set of data is released.

The HGBS observed nearly all of the star forming clouds in the solar neighbourhood, but recent work has called attention to the California-Auriga molecular cloud.  Despite being as massive as Orion ($\sim 10^5~M_{\odot}$), this cloud was long ignored owing to the minimal star formation taking place therein \citep{lada-cmc}.  Since it provides a counterpoint to the vigorous star formation taking place in the nearby Orion region, we include SPIRE observations by \citet{california_cloud} in our analysis.  This provides some ability to check whether the differences in star formation rate are associated with differences in the filamentary network.

Maps from the {\it Herschel} SPIRE instrument have a zero-point intensity offset introduced in the mapping procedure.  While the relative calibration within the maps is excellent, the absolute calibration requires correcting this constant offset.  Because of significant overlap in the bandpasses, we use the 350 $\mu$m data from the second public Planck data release  \citep{planck-dr2}.  Following \citet{bernard10}, we determine the offset by convolving the SPIRE data to match the Planck resolution, correlating the data position-by-position, and fitting a line to the correlation.  We find that the relative calibration is excellent as measured by the slope of the line: within 10\% of unity, with the differences arising in part because of different bandpass shapes between the Planck HFI 857 GHz band and the SPIRE 350 $\mu$m band.  The zero-point offset is determined from the constant offset for the line and is reported in Table \ref{tab:properties}.  We note that, because of the adaptive thresholding employed in the filament identification, most of the results are practically unaffected by this correction.

Since the flux ratio between the 350 $\mu$m band and the other SPIRE bands is a function of dust emissivity and temperature, the Planck referencing is ambiguous for other bands.  While the filament identification algorithm works well on these other images, we focus this preliminary analysis on the 350 $\mu$m data because of the clear route to absolute calibration.

\section{Results and Analysis}
\label{section:results}

In this section, we present results determined purely from the filament analysis of the 350 $\mu$m emission alone (\S\ref{section:results}).  We then examine those results in the context of preliminary physical interpretations (\S\ref{section:discussion}).  These latter analyses are not complete because dust emission is a function not only gas column density but also dust temperature, gas-to-dust ratio, and dust emissivity.  These confounding variations of dust properties preclude a direct translation between emission and column density.   Multiband analysis of filamentary structures finds that they show an increase in dust temperature with filament radius \citep[e.g.,][]{konyves-2010-aquila}, and dust emissivity shows evidence for varying with different density histories \citep{schnee05,schnee06,schnee08,schnee10}.  Other efforts are in progress to provide definitive maps of the physical conditions in these clouds and we confine our analysis to a restrictive linear scaling between dust emission and gas column density.  While this indicates our interpretations are potentially inaccurate, the main purpose of this work is to present a new methodology and indicate novel directions for robust analysis, and a full treatment of these effects is beyond the scope of the present work.
\begin{figure*}
\mbox{
\subfigure{
\includegraphics[scale=0.35, clip=true, trim=0cm 0cm 0cm 0cm]{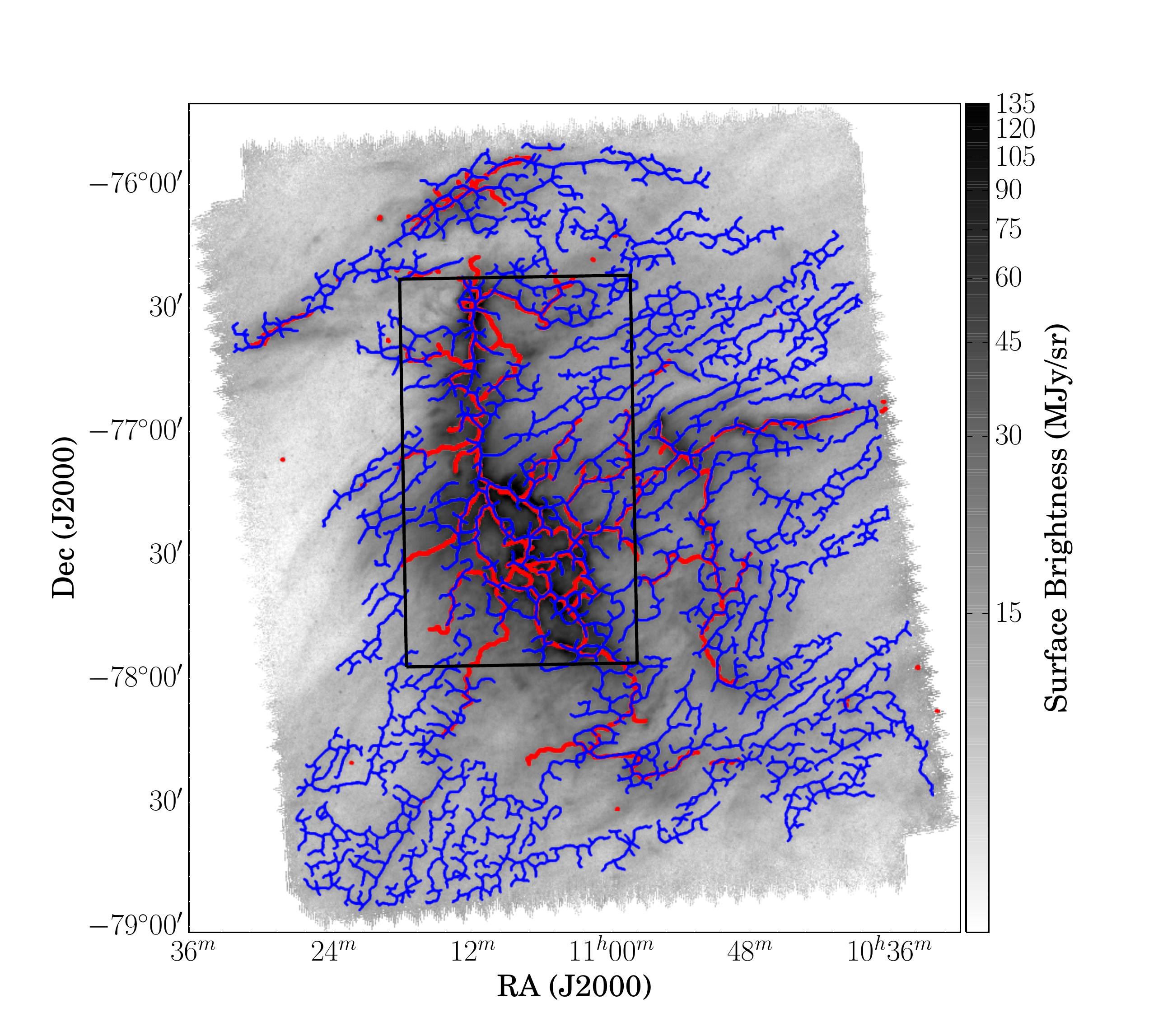}
\label{fig:disperse}
}\quad
\subfigure{
\includegraphics[scale=0.35, clip=true, trim=2cm 1cm 1cm 1cm]{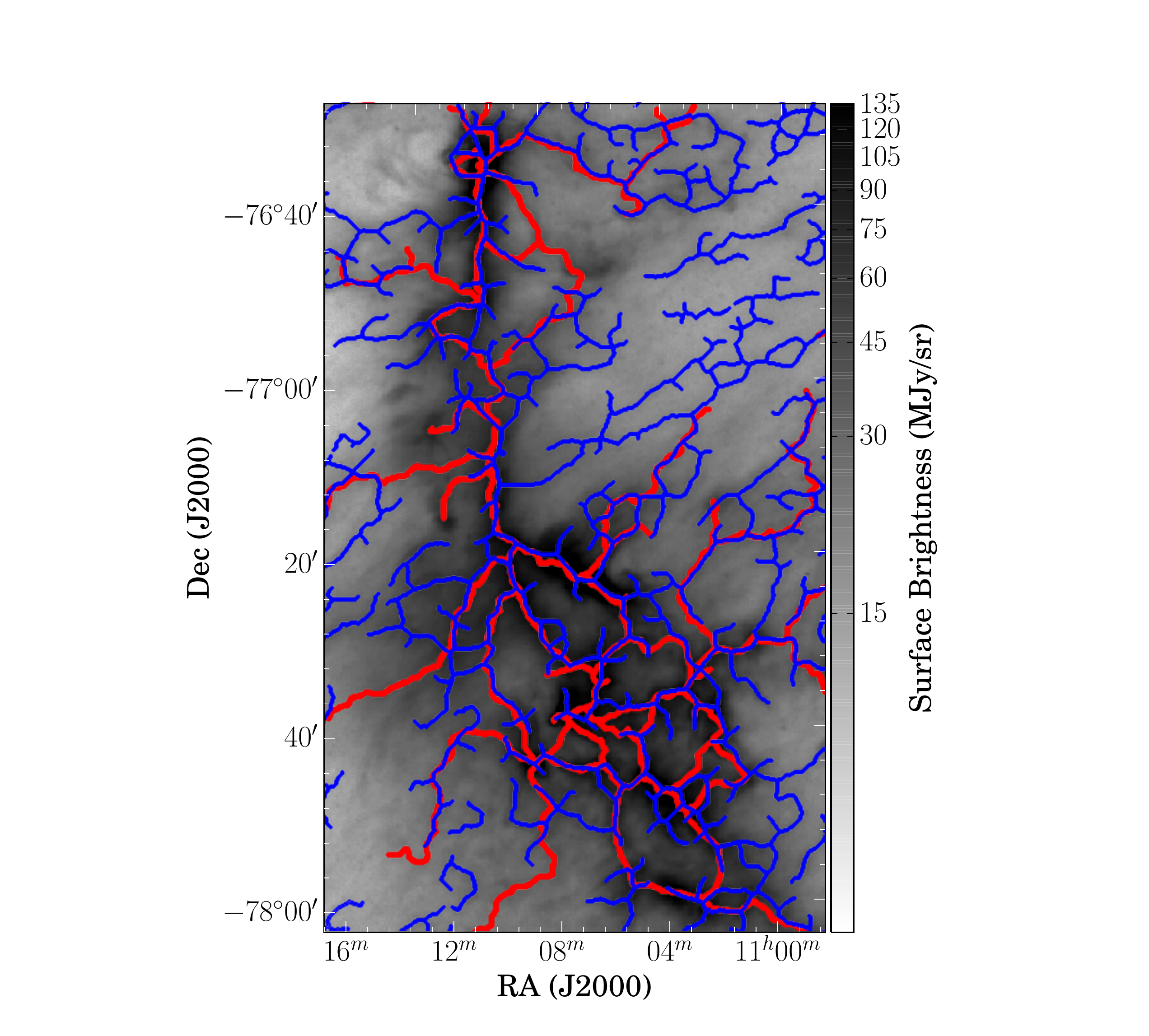}
\label{fig:disperse_zoom}
}
}
\caption{Left: A comparison of filaments returned by the DisPerSE algorithm \citep{disperse} (red) and those returned by our method (blue) in the Chamaeleon region. Right: A zoomed in view denoted by the box in the left.  In general, the filaments identified using {\it FilFinder} agree well with those identified by DisPerSE.  However, {\it FilFinder} recovers many more faint, filamentary features away from the brightest regions in the image.}
\end{figure*}

We analyze the SPIRE 350 $\mu$m data for the 14 regions described in Table \ref{tab:properties} above.  The maps are convolved using a Gaussian beam to a reach a common physical resolution set by the angular size of the $25''$ SPIRE beam at the maximum distance present in the analysis (460 pc), which projects to 0.056 pc. We also regrid the data to a common pixel scale based on the smallest distance of the clouds analyzed (140 pc). This removes any pixel dependencies, which could skew the derived properties of each region.  We then apply the filament finding algorithm as described in Section \ref{section:methods}.  An example of the results of the analysis for the Chamaeleon region are shown in Figure \ref{fig:disperse_zoom}.  For reference, we also plot the filaments identified\footnote{Following the tutorial: \url{http://www2.iap.fr/users/sousbie/web/html/index38d8.html?post/regular-grid-\%3A-filaments}} using the DisPerSE algorithm described in \citet{disperse}. We used a persistence cutoff of 12 in the figure. While lowering the persistence reveals fainter features, we found that not all of the faint features found by {\it FilFinder} were recovered by DisPerSE without also recovering spurious features. Furthermore, the returned network was largely connected into one feature despite using the network breakdown option in the package. The {\it getfilaments} code of \citet{getfilaments} is not publicly released at present.  The DisPerSE and {\it FilFinder} algorithms reproduce the general shapes of the brightest filamentary structures.  The exact paths traced by the two approaches are not necessarily identical, but the basic features are well reproduced.  In addition, {\it FilFinder} identifies numerous features in the faint regions of emission that are undetected using DisPerSE.  These features agree well with the faint structures in the image, particularly those that have been attributed to magnetic features in clouds \citep{li-2013,rollinhough}.

\subsection{The Filament Population}
\label{sub:population}

\begin{figure*}
\includegraphics[width=0.9\textwidth]{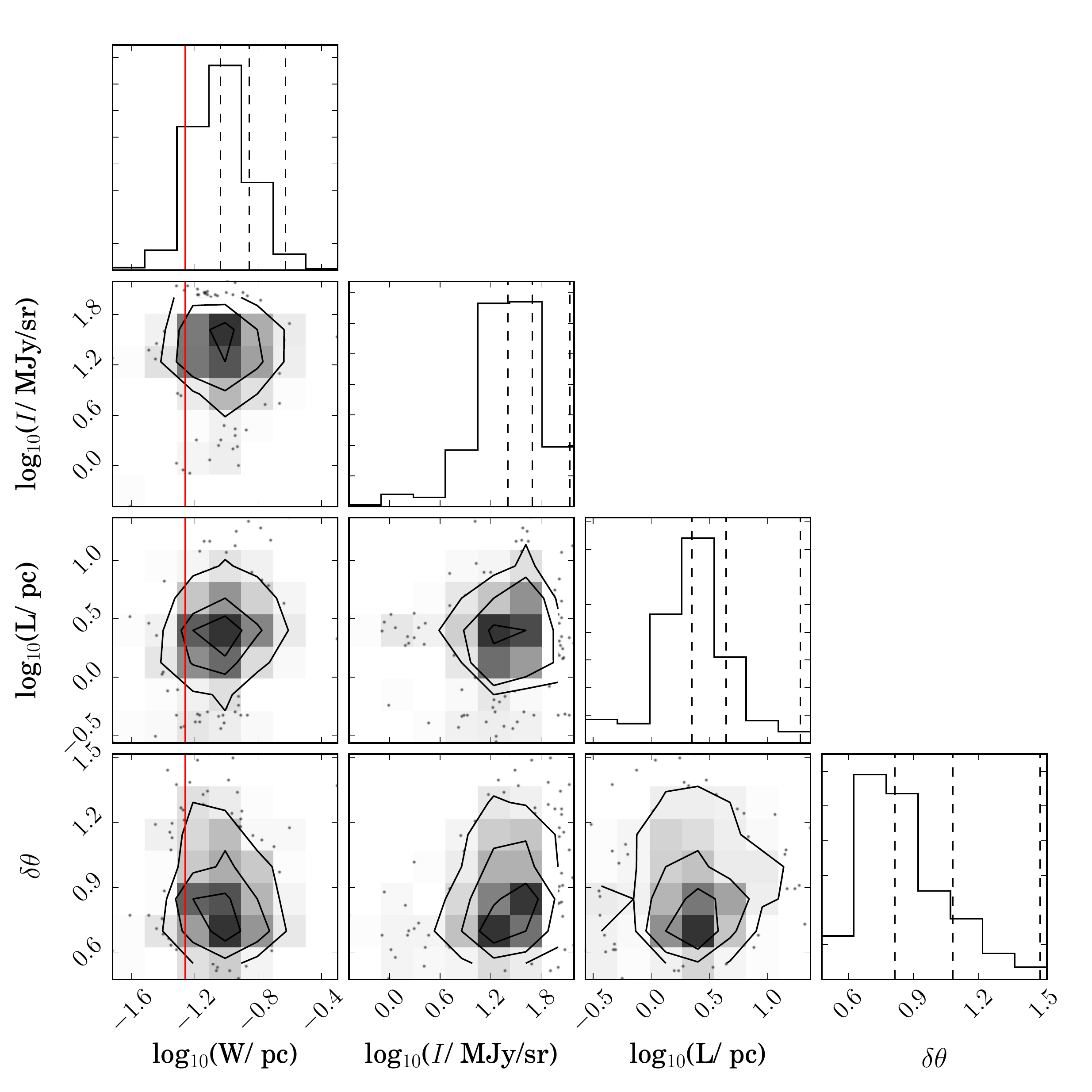}
\caption{\label{fig:fil_props} Key properties derived from for all filaments in the HGBS and the California Molecular Cloud \citep{california_cloud}. Note that the width refers to the deconvolved FWHM. The dotted lines indicate the 50, 85 and 99.5 percentiles of the histograms. The contours represent the areas containing the 50, 85 and 99.5 percentiles. Dots represent individual data points that do not fall within the lowest contour. All filaments are above the minimum length cutoff (5 times the beamwidth, or 0.275 pc for the common projection to a distance of 460 pc), and so it is not shown in the figure. The beam width of 0.056 pc (projected at 460 pc) is indicated by the red line.}
\end{figure*}

In Figure \ref{fig:fil_props} we present the joint distributions of filament properties for all the regions described in Table \ref{tab:properties}.  We adopt notation for filament length ($L$), width ($W$), curvature ($\delta\theta$), orientation ($\theta$), average surface brightness ($I$).  In Appendix \ref{app:parameter_sensitivity}, we also argue that our parameter distributions are unlikely to be a significantly affected by the resolution of the 350 $\mu$m data.  Any small effects should be removed by the convolution of all data to a common physical resolution and projection to a common pixel scale.  Of the total 369 filaments found, the radial profile fits failed for just 11 filaments due to the complications discussed in Section \ref{sub:width}. In each of these 11 cases, the width was too small to be deconvolved; these are the only filaments with widths that could not be deconvolved. Overall, we find properties similar to previous analyses. Filament brightnesses range mostly between 20 and 70 MJy~sr$^{-1}$, and are typically around 25 MJy~sr$^{-1}$. These values correspond to the peak brightness above the background in the fit. The median deconvolved filament width (FWHM) is 0.09 pc with some spread around this width (0.06 and 0.14 pc at the 15th and 85th percentiles, respectively), in agreement with \citet{pp6-andre}. There is no significant covariance with width: bright filaments show no systematic difference in width compared to fainter filaments.  Filament lengths are typically 2.2 pc long with the majority ranging between 1.4 and 4.4 pc. In exceptional cases, a small number of filaments reach up to 20 pc.  Note that filament lengths are path lengths and the projected distance between the ends is necessarily smaller. Furthermore, this length does not take into account projection effects due to the unknown inclination. Based on the ranges of lengths and widths, the typical filament aspect ratio is between 7 and 9.

\subsection{Effect of Cores on Filament Properties} 
\label{sub:effect_of_cores_on_filament_properties}

As indicated in Section \ref{section:methods}, we derive filament properties based on the original images and do not attempt preprocessing to remove cores. In this section, we briefly describe how the inclusion of cores affects the properties of the filaments. As noted in several studies of the HGBS  \citep{konyves-2010-aquila,arzou2011_ic5146,pp6-andre}, the width of cores corresponds well to previously derived widths of filaments ($\sim$0.1 pc). Based on a by-eye analysis of our results, we find that the skeletons returned by {\it FilFinder} correspond well to tracing the centre of the cores along the filament. We note, however, that the skeletonization process used does not guarantee this, leading to the possibility that the radial profile is artificially wider due to the offset of a bright feature. Very few of these cases were found upon inspecting the radial profiles of each region. We expect that this has a negligible effect on the global properties derived.

Filaments with significantly higher surface brightness have been found to have many cores along their extent \citep[e.g.,][]{konyves-2010-aquila}. This large number of cores likely causes a significant increase in the peak of the filament's radial profile. We expect that this is the greatest effect of including the core population while deriving filament properties. Since the number of low surface brightness filaments far exceeds those of high surface brightness, this effect will inflate the positive tail of the surface brightness distribution (Figure \ref{fig:fil_props}).

Finally, we note the dependency of both of these potential issues on the size of the filament. Many filaments have skeletons which cover regions over a large dynamic range in surface brightness. Often the portions of lower surface brightness extend substantially from the brighter regions. In these cases, both of the issues presented above are unlikely to have a large effect on the radial profile due to the large number of pixels at lower surface brightness.   Future work that includes a joint analysis of compact core objects and filaments will mitigate these possible biases.


\subsection{Region-based Analysis}
\label{sub:region_based_analysis}

The preceding section aggregates all the regions together into a single analysis, but this combines widely different regions of the ISM.  We take advantage of this broad survey of the Gould Belt region to compare the distributions of properties in different regions.

\begin{figure*}
\includegraphics[width=0.9\textwidth]{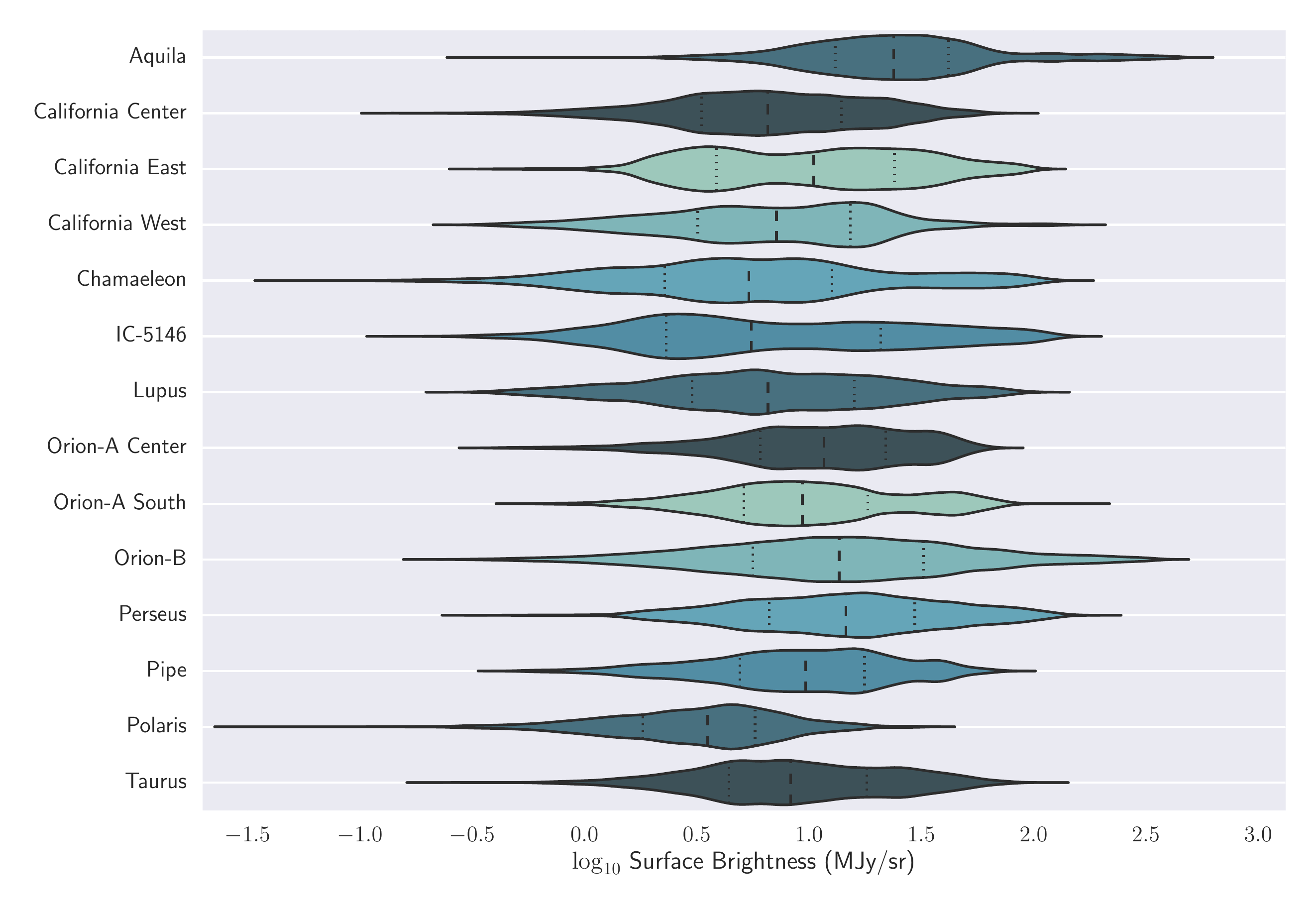}
\caption{\label{fig:violin_plot} A violin plot showing the background subtracted surface brightness along the filamentary structure.  The `violin' shows an estimate of the PDF for the surface brightness.  These regions show significant multimodal structure in their brightness PDFs and significant variations between regions.}
\end{figure*}

In a by-eye inspection, the brightness of the filamentary networks appears to vary significantly cloud-by-cloud (and in regions across the face of the cloud).  We summarize these variations by plotting the distributions of filament brightness above the background level in Figure \ref{fig:violin_plot}.  This violin plot shows amplitude of the probability density functions (PDFs) as the full width of the `violin.'  The probability density function is generated from a Gaussian kernel density estimate of the observed data, which smooths the PDF.  The quartiles of the data are shown as lines drawn in the distributions.  This visualization is similar to stacked histograms of the data but highlights the differences in shape between the regions.  Few of the PDFs are Gaussian and many (notably Orion A South, California East and IC-5146) show bimodal structure with the bright mode an order of magnitude brighter than the faint mode.  Next, the brightness distributions, while overlapping, are significantly different between the regions.  Aquila is dominated by relatively bright filaments whereas the Polaris region is substantially fainter.  The California molecular cloud shares a common mean and width for the distribution across all images analyzed for this cloud, but the distribution shape varies over the cloud.  Similarly, the two parts of Orion A share a common distribution, though the ``Centre'' of the cloud which contains the main ridge of the cloud shows no bright mode, indicating that the southern portion of the cloud has more mass in the filamentary structure.   The Orion B PDF spans the range seen in the Orion A images, but still sits at larger values than the lowest mass clouds.

We estimate the influence of filamentary structure on the regions as a whole using two different measurements.  First, we compare the typical filament brightness to the background level estimated in the width. On average, filament brightnesses are 1.3$_{-0.2}^{+0.5}$ times that of the broad background on which they are found. This factor does not show significant variation from cloud to cloud across the Gould Belt.  We also use our model to construct an image of the filamentary network to quantify what fraction of cloud emission in the entire image is found within the filamentary network. We find typical values between 4.5\% and 23.9\% with no clear relation to cloud mass or star formation properties. Note that this quantification includes all the emission in the image, which may spuriously include foreground and background emission not associated with the cloud. These values reflect only the emission contained within the width of the filament above the background level derived in the fits; thus emission on scales larger than the filament width is not included. These values indicate that the filamentary network, while prominent describes a relatively a small fraction of the total emission in the cloud.

\begin{figure}
\includegraphics[width=0.5\textwidth]{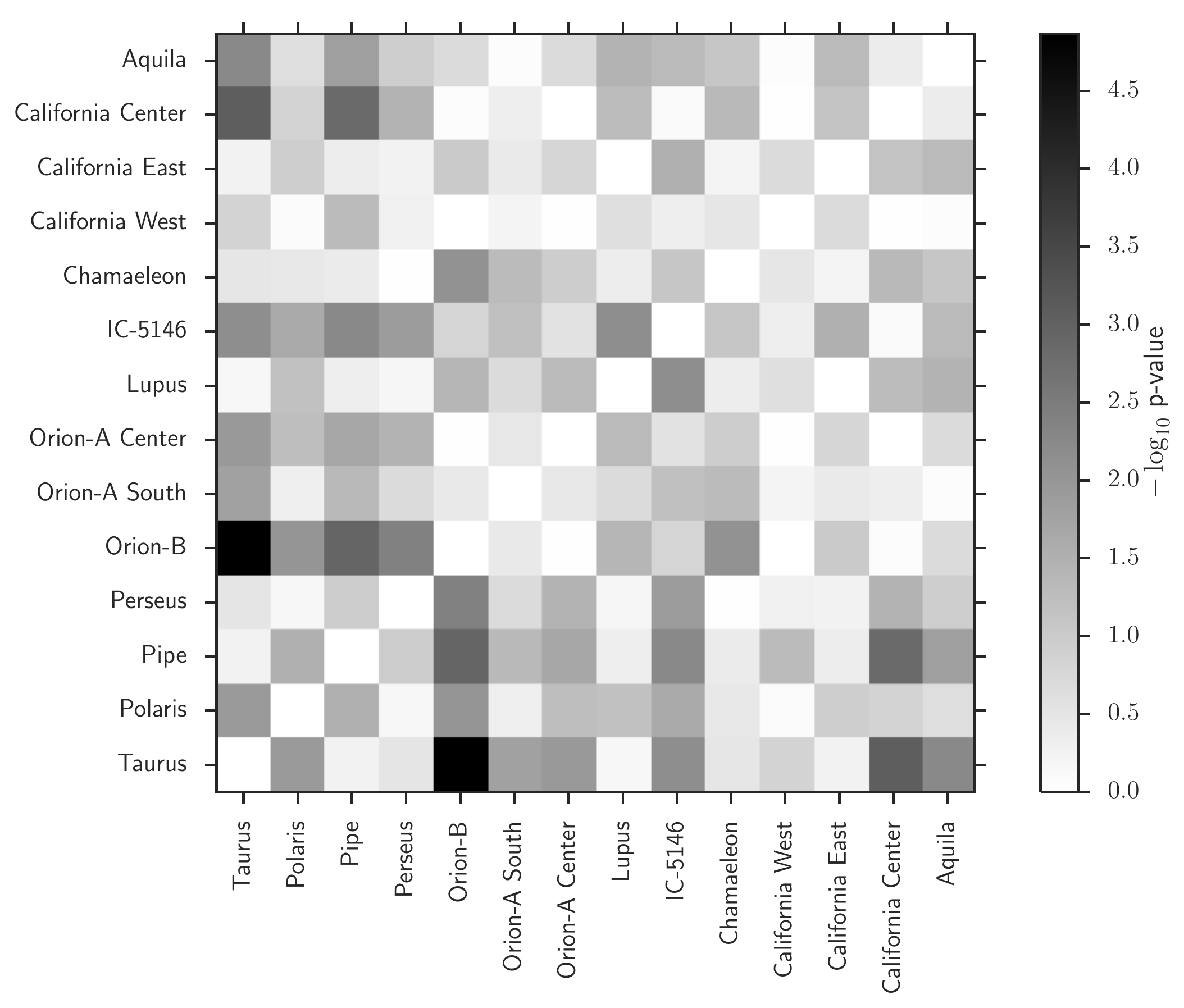}
\caption{\label{fig:ks_tests} P-values for two-sided Kolmogorov-Smirnoff tests showing the probability of the filament distributions being drawn from the same parent distribution for the filament width. The greyscale value represents the negative logarithm of the probability so substantially different distributions are larger (darker).}
\end{figure}

We compare the width distributions for the filament families using a two-sided Kolomorov-Smirnoff test \citep{nr}.  In Figure \ref{fig:ks_tests} we show the negative log-probability that the two distributions are drawn from the same parent distribution.  The Figure indicates that there are detectable variations in the filament width distributions between clouds.  The clearest associations between populations suggests some variation with either cloud mass or distance, since the GMCs are, on average, more distant than the low mass clouds.  The algorithm shows good robustness with respect to distance (see Appendix \ref{app:resolution}) and we carry out all the analysis with data convolved to a common physical resolution.  However, there may be some distance effects introduced because the more distant objects are physically larger in a given image and may sample a wider range of properties.

Overall, we find good evidence for region-based variation among the different nearby star-forming molecular clouds.  Several of the {\it Herschel} maps observe the same object but have divided the map into different regions based on observing requirements.  In general, we find little variation between images observing different parts of the same object.

\subsection{Filament Orientation}
\label{sub:orientation}

\begin{figure}
\includegraphics[width=0.48\textwidth]{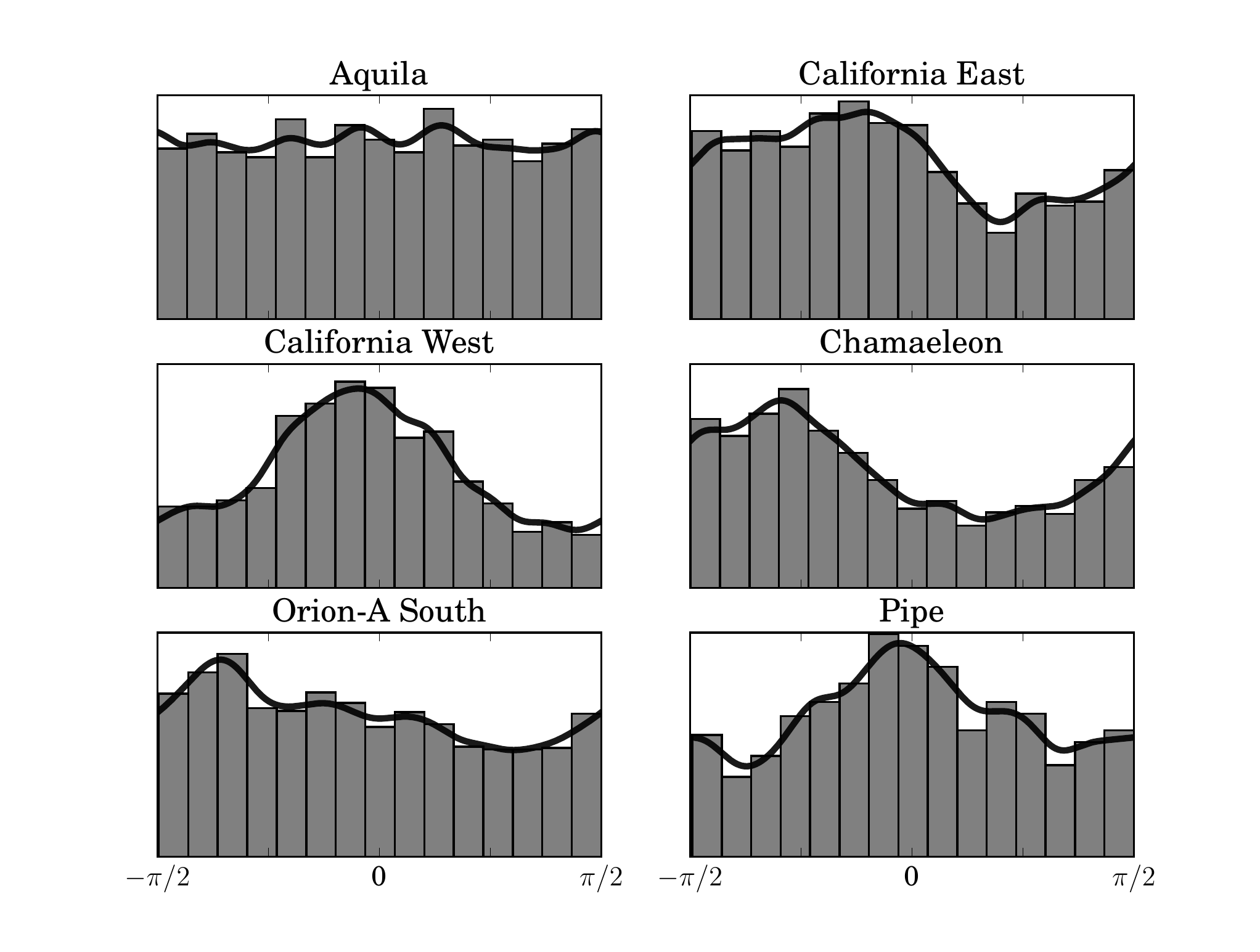}

\caption{\label{fig:orientations} Filament orientations for six Gould Belt regions. All histograms are normalized to unity and the kernel density estimations (solid black line) are computed over a continuous boundary.  The East-West direction corresponds to 0$^{\circ}$.  Some clouds show preferred orientations (California West, Pipe, Chamaeleon) and others are uniform in the brightness distributions (Aquila). }
\end{figure}

The Rolling Hough Transform (RHT, \S\ref{sub:plane_orientation_and_curvature}) provides a novel means to assess the orientation of linear features.  In our case, we use it to search for preferred orientations in the filamentary networks that may be caused by magnetic fields or the directions of converging flows.  We search for the potentially related signatures of the faint striation populations relative to brighter filaments. Here we adopt the nomenclature that a `striation' is a faint filament. Despite the fact that some clouds show bimodality between bright and faint features, we stress that our analysis in Section \ref{sub:population} does not show evidence for a universal intensity threshold value that separates the filaments into bright and faint populations. As such we do not specify a threshold to indicate filaments that can be definitively considered striations. We show the orientation distributions for six Gould belt regions in Figure \ref{fig:orientations} that exhibit the range of orientation distributions in the analyzed clouds.  We measure the angular distribution of the sets of branches that comprise the filaments rather than the distribution for entire skeletons as a whole.  While this destroys orientation information for individual filaments, this approach better describes the orientation of the region's entire filamentary structure.

{\it Chamaeleon} -- \citet{chamaeleon-striations} noted that the Chamaeleon Molecular cloud shows two distinct filament and striation populations  (see Figure \ref{fig:disperse}). By eye examination of the images suggests that the populations intersect each other perpendicularly.  The RHT results show a significant peak in the direction of the striation population which dominates the region by surface area.  These striations lie in the direction of the large scale magnetic field \citep{chamaeleon-striations}.  A small peak can be seen approximately in the direction of the bright filaments. Qualitatively, the Chamaeleon region has a bimodal distribution of the filamentary structure orientation.  We attempted to capture this behaviour quantitatively through different brightness weighting schemes but were not able to improve on this qualitative result.

{\it Aquila} -- The Aquila region shows no preferred direction for the filamentary structure. This is consistent with a by-eye evaluation of the data. As a whole, the region shows no clear bimodality as we see in Chamaeleon.

{\it California} -- Despite being part of the same complex, the East and West parts of the California Molecular Cloud show some differences in the orientation of the filamentary structure. The West component has an aligned striation structure, though not clearly perpendicular to the prominent filaments as seen in Chamaeleon. This is reflected in the output of the RHT (Figure \ref{fig:orientations}), which shows a prominent peak centred near 0 radians. There is no noticeable peak corresponding to the bright filaments. The central region of California (not shown) reveals a similar RHT output as the West component. Conversely, the East component yields a more uniform distribution with a wider peak centred near $-\pi/4$. The uniformity makes it more akin to the Aquila region than the other regions shown in Figure \ref{fig:orientations}. The striation structure in this region may have been disrupted by the concentrations of YSOs in the region \citep{california_cloud}.

{\it Pipe} -- The Pipe molecular cloud has a unique structure possibly attributed to a shock \citep{pipe-filaments}. There is a clear directional preference centred at $\theta=0$, consistent with the shape of the bow-front. The small peak corresponds to the direction of the connecting structures located behind the bow-front. Unlike other regions where the faint filamentary structure dominates the orientation information, the Pipe region is dominated by the prominent filaments. There is a noticeable lack of a striation population that is found in most other regions in the Gould Belt. This absence may be further evidence of the bow-shock dominating the region.

{\it Orion-A South} -- The south region of Orion-A has a preferred directions at about $\theta=-1.4$ with less significant peak closer to $\theta=-0.4$. In comparison with Chamaeleon, these peaks are less significant and the distribution is more uniform. The peak at $-0.4$ radians corresponds approximately to the direction of the bright filaments in the region. The peak at $-1.4$ radians is associated with the striations, particularly near the northern part of the image.

The orientation histograms presented in Figure \ref{fig:orientations} are not weighted by an intensity value associated with each branch. We find that such a weighting scheme causes no significant change in the histogram's shape for all regions.  Since intensity scales with filament mass, this points to a clear indication that the faint striation structure as a whole contributes a significant fraction of the mass.  The orientation of these striations is likely connected to the large-scale magnetic field around the clouds.  In addition to the work of  \citet{chamaeleon-striations} in Chamaeleon, the alignment of the faint linear features in the ISM with the large scale magnetic field has also been noted in dust extinction \citep{li-2013} and {\sc Hi} \citep{rollinhough}.  The sensitivity of the {\it FilFinder} algorithm to faint filamentary structure offers a method to quantitatively identify the signature of the magnetic field.

\section{Discussion}
\label{section:discussion}

Examination of the filament populations showed significant region-to-region variation between the different sections of the Gould Belt.  Since such clouds also show significant variation in their star formation activity, we explore some empirical correlations between filaments and star formation.

\subsection{Star Formation and Filamentary Structure}
\label{sub:sfr}

The scientific interest in filaments has largely been driven by their close connection to the star formation process.  Since this work presents a means to identify filaments across a range of molecular clouds, we explored whether the properties of the filamentary network were connected to the star formation rates (SFRs) of clouds.  We have adopted the SFRs for these regions described in the literature (see Table \ref{tab:properties}) and parameterize them in terms of a SFR surface density: $\Sigma_{\mathrm{SFR}}$.  In general, these studies identify the numbers of protostars of different classes, and assuming an evolutionary timescale for the different classes, infer a SFR.  These rates are not spatially matched to the SPIRE images we analyze in this study.  In cases where a single SFR is supplied in the literature for a region, we apply the rate to all images associated with that region.

\begin{figure}
\includegraphics[width=0.48\textwidth]{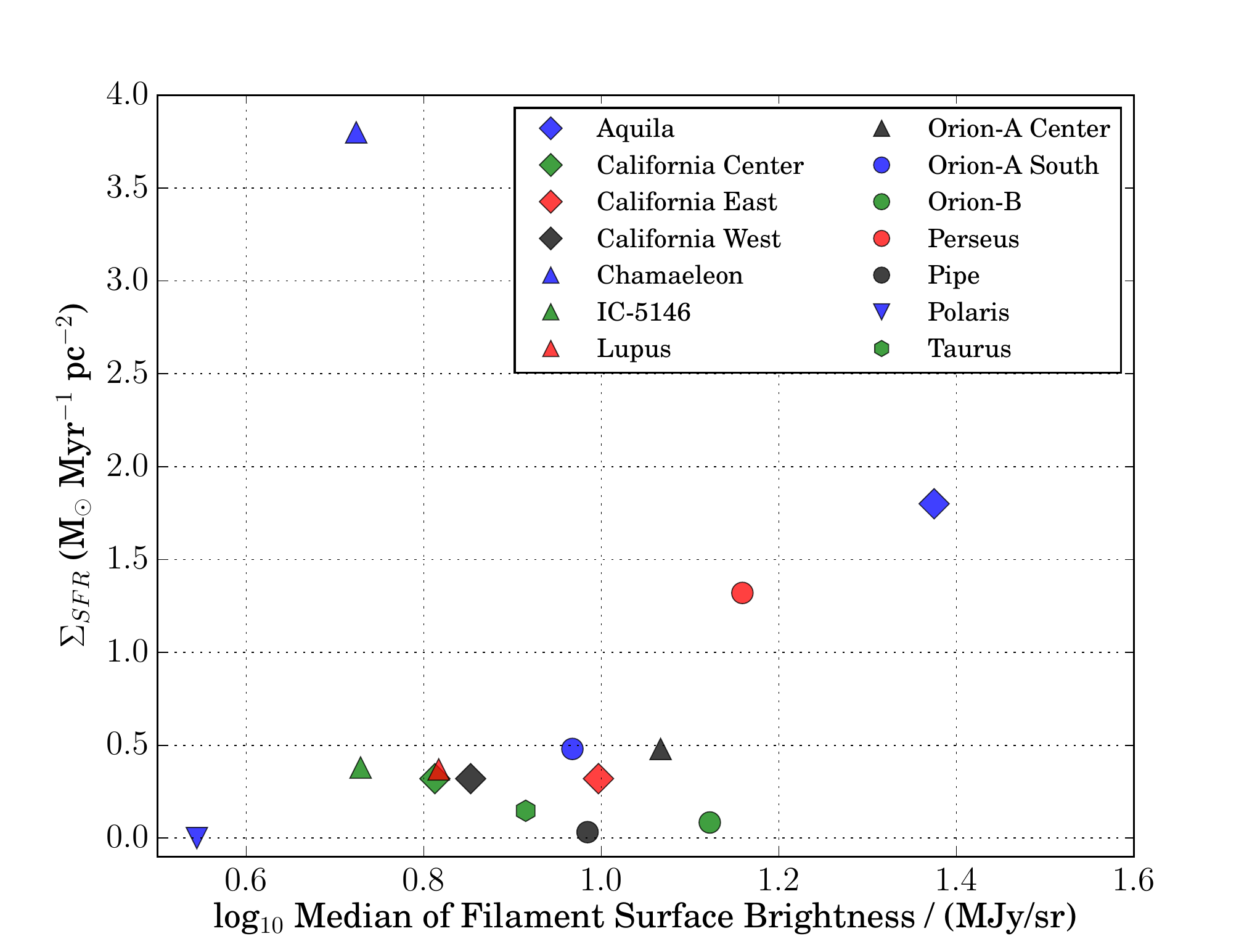}
\caption{\label{fig:sfr_vs_bright} The SFR densities (see Table \ref{tab:properties}) plotted against the median of the surface brightness along the filamentary structure.   There is a suggestive trend of the surface density of star formation with typical filament brightness though the Chamaeleon region is an obvious outlier.}
\end{figure}

In Figure \ref{fig:sfr_vs_bright}, we plot $\Sigma_{\mathrm{SFR}}$ against the median filament brightness for the different images.  This parameterization explores connection between the strength of the filamentary structure in a region and star formation rate.  The {\it FilFinder} approach explicitly includes a smooth background of emission underneath the filamentary network, so the analysis specifically measures the magnitude of the filamentary substructure rather than the cloud surface density overall.   There is not a clear correlation given the available data; however, with the exception of Chamaeleon, images with brighter filamentary structure tend to have higher $\Sigma_{\mathrm{SFR}}$.  The Chamaeleon region is small and may have its star formation rate estimates affected by small number statistics.   The weak correlation is driven by Aquila, Perseus and Orion A all having higher than average filament brightnesses and $\Sigma_{\mathrm{SFR}}$.

Since the algorithm also fits a background level, above which the filament brightness is measured, we compared the filament brightness to the background.  This background emission could be either the relatively smooth bulk of the cloud or true background emission, unassociated with the star forming region but projected along the line of sight.   We find that all clouds have a typical filament brightness that is typically 1.3 times larger than the background emission on which the filaments are found. The bright filaments can have filament-to-background ratios of more than 10.

We also explored the correlation of other properties of the filamentary network with star formation rates but did not find any obvious connections.  However, these results may become clearer with the full release of HGBS data including column density and temperature results.  If borne out by a more thorough analysis, this type of correlation would support the evolutionary sequence where filaments form and gravitationally fragment \citep{pp6-andre}, possibly mediated by magnetic fields (Chen \& Ostriker, in preparation).  In this scenario, the early stages of molecular cloud evolution would be parameterized by the amplitude of the filamentary network within the cloud.  Our method provides a method to characterize this network.

\subsection{Filament Stability}
\label{sub:stability}

\begin{figure*}
\includegraphics[width=0.9\textwidth]{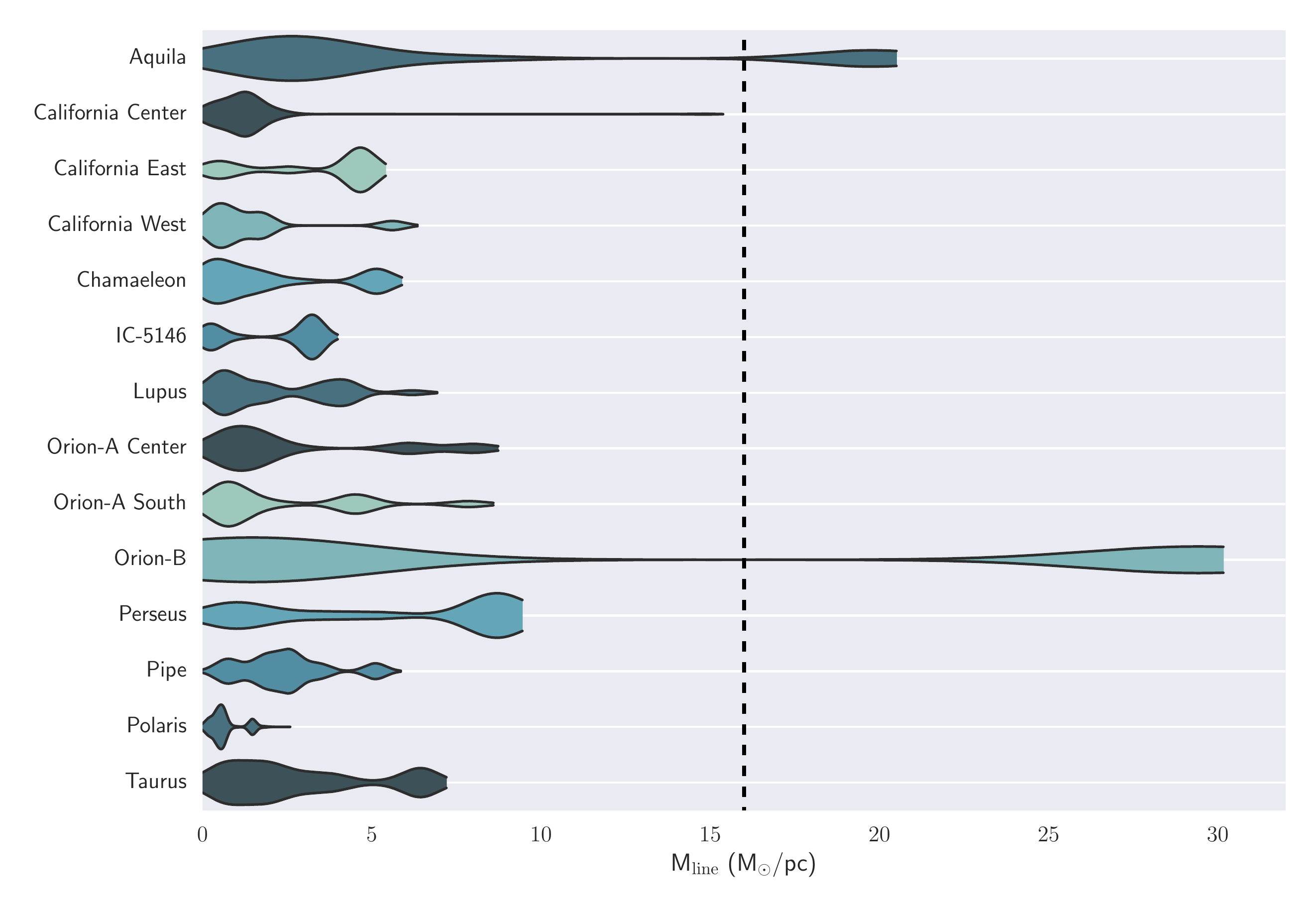}
\caption{\label{fig:violin_Mline} Average linear mass densities of the filaments for a constant $T_{\mathrm{dust}}=15$~K. The density is calculated by including all pixels within the FWHM of the filament. The distributions in the plot are weighted by the number of pixels in the filaments. The dotted line corresponds to the thermal critical mass density of 16 M$_{\odot}$~pc$^{-1}$}.
\end{figure*}

We also explore the distribution of filamentary emission that would be subject to classical gravitational instabilities.  For this initial analysis, we adopt a constant dust emissivity at 350 $\mu$m of 0.073 cm$^{2}$~g$^{-1}$ from the law of \citet{beckwith90} and the assumption that dependence on frequency scales with the index $\beta=2$.  We adopt a uniform dust temperature of $T_{\mathrm{dust}}=15$~K as is characteristic of the more careful analysis by \citet{konyves-2010-aquila} and \citet{arzou2011_ic5146}.  Given these assumptions  with gas-to-dust mass ratio of 100-to-1, we find $\Sigma_{\mathrm{gas}}=1.02~M_{\odot}\mbox{ pc}^{-2} (I_{350}/\mathrm{MJy~sr}^{-1})$.  Work from \citet{konyves-2010-aquila} shows that the dust temperature decreases toward the centres of filaments, meaning that the values derived here for the mass will be lower limits on the true surface density.  Using the modified blackbody model for the spectrum, this difference would be a factor of 4 for $T_{\mathrm{dust}}=10$~K.

The critical mass per unit length for a nearly isothermal cylinder is $M_{\mathrm{line,crit}}=2c_s^2/G$ \citep{inutsuka97}.  Taking $T_{\mathrm{gas}}=10$~K as typical for a molecular cloud and the mean particle mass in molecular material to be $\mu = 2.4$, we find $M_{\mathrm{line,crit}}=16~M_{\odot}~\mathrm{pc}^{-2}$.   We calculate the observed $M_{\mathrm{line}}$ assuming that the filament has a Gaussian profile perpendicular to the filament with a FWHM as measured previously: $W$.  We average the measured surface densities over the filament within the FWHM width of the filament: $\langle \Sigma_{\mathrm{gas}}\rangle$ and compute the linear mass density as $M_{\mathrm{line}}=\langle \Sigma_{\mathrm{gas}}\rangle W / \mbox{erf}(\sqrt{\ln 2}) = 1.23 \langle \Sigma_{\mathrm{gas}}\rangle W$.  The numerical prefactor arises because of the assumed Gaussian profile and the correction required to account for the wings of the Gaussian outside of our integration region.  More careful treatment of projection effects and direct measurements of $c_s$ will be required to assess whether a given filament will be subject to gravitational instability.  Moreover, the instability criterion ignores other forms of support such as turbulence or magnetic field configurations that would make this threshold value a lower limit.

\citet[][in preparation]{chen-ostriker-aas} argue that the presence of velocity gradients across filaments seen in the CARMA Large Area Star formation Survey data \citep{classy-b1} indicates that filament formation occurs in a magnetized slab.  In such an environment, the magnetic critical length is $L = B/(2\pi \rho G^{1/2})$.  Taking the filaments as having an axisymmetric Gaussian density profile gives $M_{\mathrm{line}}\sim \Sigma L \sim B W /(2\pi G^{1/2})\sim 9~M_{\odot}\mbox{ pc}^{-1}$.  Here we have taken the typical width of a filament to be 0.1 pc consistent with our analysis and the typical magnetic field to be $B=30~\mu\mathrm{G}$ as a summary value for the results from the \citet{planck-local-bfield}.  This value is of order the gravitational instability value, but depends on local field strength and filament width.  These parameters can be related back to mass densities using relationships between pre- and post-shock values \citep{chen-ostriker-aas} but this relationship requires assumptions about the strength of the shock.  While notable that the different instability mechanisms produce similar linear mass densities, the measurements that ground the theory of filament formation via magnetohydrodynamic processes are difficult to obtain from these data alone.  This estimate suggests that the surface densities required for gravitational collapse will also be able to overcome magnetic support as well.

In Figure \ref{fig:violin_Mline}, we show a violin plot for the linear mass density of filaments in the different clouds analyzed here.  Overall, the linear mass densities of clouds are below the stability criterion, but the massive star forming clouds show bright filaments well above the instability threshold.  While the comparison to the values required for instability is interesting, several factors preclude strong conclusions.  Some filaments in the other regions may be unstable since lower-than-assumed dust temperatures can significantly increase the mass densities (by a factor likely between 1 and 4).  Further, the filaments show significant curvature along their lengths and this `kinking' would make the systems more unstable.  We conclude that many filamentary networks are near the threshold of instability to fragmentation.  Massive molecular clouds such as Aquila, California (Centre) and Orion show a few bright filaments that would be unstable over many different combinations of assumptions, but the lower mass clouds do not have the same significant bright filaments.

\subsection{Future Work}

This paper presents a first look at filamentary structure through the lens of a new algorithm.  While some of the results show that the new method is promising, future work is required to make the outcomes more robust.  Using robust maps of the column density and temperature derived from multi-band fitting is the first obvious improvement, potentially with links to the {\it Planck} data on the largest angular scales.  With such maps, we expect an increase in the mass densities of the filaments and possibly a decrease in width as the centres of the filaments gain more importance in the fits.  The curvature of the filaments and their lengths are unlikely to be significantly altered with more physically grounded data.

A primary advantage of this approach, shared with the DisPerSE algorithm, is the ability to extend naturally to three dimensions.  This extension would allow for filament extraction from position-position-velocity (PPV) data cubes of spectral line emission.  The primary barriers to this innovation lie in the careful treatment of intersection and pruning for three-dimensional skeletons.  The algorithm is stabilized by the excellent signal to noise present in the {\it Herschel} data.  Typical millimetre-wave PPV data sets have significantly poorer sensitivity but new ALMA data sets will have the sensitivity required to stabilize the algorithm.

Finally, a joint analysis of core and filament populations will eliminate some of the concerns that cores are distorting filament properties (\S\ref{sub:effect_of_cores_on_filament_properties}).  Furthermore, including background filament structure in a joint analysis with the cores will improve estimates of core size, stability and connection with their background environment.  With the code publicly released, we welcome community contribution to its development.

\section{Summary}

We have presented {\it FilFinder} a new algorithm for measuring filaments in astronomical images. {\it FilFinder} adopts the techniques of mathematical morphology to identify filamentary features in image data.  This approach differs from other algorithms, which use wavelets and curvelets ({\it get\_filaments}) or critical manifolds (DisPerSE) or Hessian operators \citep[the Hi-GAL survey and][]{salji15}. The algorithm relies on adaptive thresholding, which creates a mask based on local changes in brightness. The Medial Axis Transform reduces the signal mask to a skeleton.  This skeleton is then pruned down to a filamentary network.  When compared to the filaments identified by DisPerSE, {\it FilFinder} identifies the same bright filaments but also reliably extracts a population of faint filaments (striations) in the peripheries of the molecular clouds.

Such a morphology-based approach enables new types of analyses.  We apply this method to study the properties of the filamentary networks in molecular clouds found in the Gould Belt star forming region.  We use the {\it Herschel} Gould Belt Survey data of \citet{gbs-andre} supplemented by data on the California molecular cloud from \citet{california_cloud}.  These data provide 350 $\mu$m imaging of these clouds at a linear resolution of $<0.056$ pc.  We compare the filament populations between different molecular clouds in the Gould Belt.  Because of the preliminary nature of the data released by the {\it Herschel} Gould Belt Survey team, our results are tentative and serve to illustrate the types of analysis enabled by the new approach.  In this preliminary context, we find:
\begin{enumerate}
\item Filament widths are found here to have a median value of $W=0.09$ pc across the cloud surveyed.  The distribution shows a variation in width of a factor of $\sim 2$.  The striations extracted here alongside the usual bright filaments have a similar distribution of widths.
\item The brightness of a filament is typically 1.3 times larger than the local background emission.  However, because filaments are not pervasive, the network comprises about 10\% of the intensity of the entire image.
\item The brightness distributions of the filamentary network vary significantly from region to region.  The large molecular clouds in the sample (Orion A and B, California) tend to have brighter filamentary networks than fainter clouds.  Regions without significant star formation (Polaris) show significantly lower median filament brightness.  Many of the distributions are bimodal with a bright mode an order of magnitude larger than the faint mode.  The distributions of filament curvature and width also show significant variations across the sample.
\item Adapting the Rolling Hough Transform approach of \citet{rollinhough}, we have examined the distribution of orientations for the filaments.  Since the faint filaments dominate this analysis by number, we can identify a preferred direction in the striation population for some of the clouds (California, Chamaeleon, and the Pipe).  There is weak evidence for a perpendicular population from the bright filaments in Chamaeleon, consistent with the by-eye analysis of \citet{chamaeleon-striations}.
\item We find a suggestive but noisy correlation between the brightness of the filamentary network and the literature estimates of the star formation rate in a region.  The Chamaeleon region does not agree with this suggestive trend, showing a high star formation rate with respect to its filament brightness.
\item We examined the stability of the filaments under a linear mapping from intensity to column density.  The bright filaments in the clouds are close to or above the threshold for instability.  However, the unmeasured quantities in the instability criteria and our simple linear mapping surface density make both the data and the threshold uncertain by a factor of two.
\end{enumerate}
While such results will need to be re-examined in the context of final data from the {\it Herschel} Gould Belt Survey, they offer some suggestive directions for the analysis of filaments in the context of far infrared emission data.  The algorithm can readily be generalized to higher dimensional data, enabling a parallel analysis in high-sensitivity position-position-velocity data.  The code and documentation have been publicly released.

\section*{Acknowledgments}

This work has been supported by a Discovery Grant and a Canada Graduate Scholarship from the Natural Sciences and Engineering Research Council Canada.   This research has made use of data from the Herschel Gould Belt survey (HGBS) project (\url{http://gouldbelt-herschel.cea.fr}). The HGBS is a Herschel Key Programme jointly carried out by SPIRE Specialist Astronomy Group 3 (SAG 3), scientists of several institutes in the PACS Consortium (CEA Saclay, INAF-IFSI Rome and INAF-Arcetri, KU Leuven, MPIA Heidelberg), and scientists of the Herschel Science Centre (HSC).  We thank Neal Evans, who provided useful guidance on star formation rates in this sample. We acknowledge the anonymous referee for their thorough review that improved the clarity and presentation of the paper.  We are grateful for the open source development community that has supported this work, in particular, the work of the \citet[][{\it astropy\rm}]{astropy} and \citet[][{\it scikit-image\rm}]{scikit-image}. This research has made use of NASA's Astrophysics Data System, the NASA Extragalactic Database, and the SIMBAD database, operated at CDS, Strasbourg, France.

\bibliographystyle{mn2e}
\bibliography{refs}

\appendix

\section{Parameter Sensitivity} 
\label{app:parameter_sensitivity}

\begin{figure*}
\includegraphics[width=0.85\textwidth]{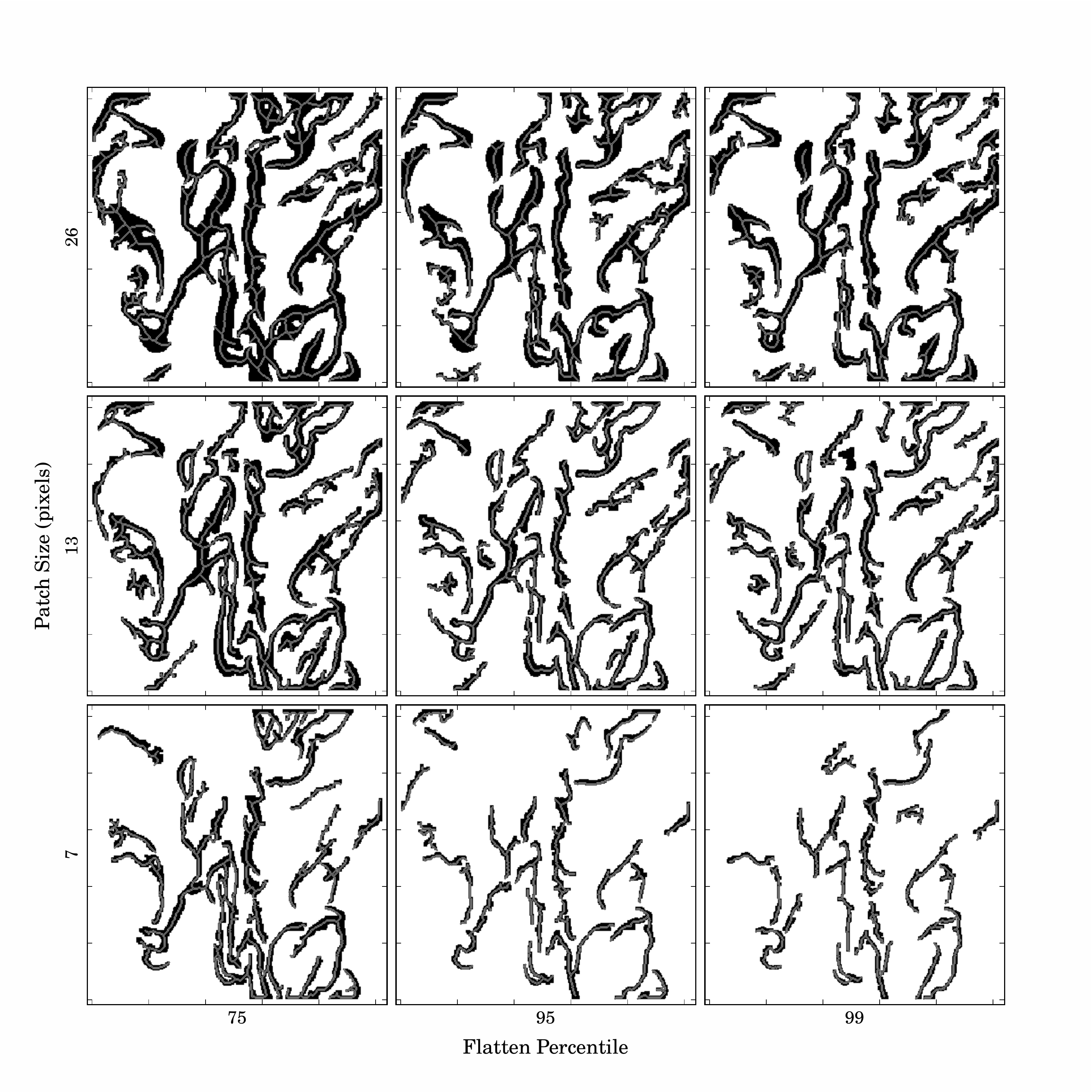}
\caption{\label{fig:patch_vs_thresh} The filament mask and cleaned skeletons from the example image shown in Figure \ref{fig:steps} are shown with varying patch size and the flattening threshold used in the arctan transform. Both parameters have a significant effect on the connectivity of the network. Significant features missing in some panels have split into smaller regions, which are rejected by the described criteria for a good filament (see \S\ref{section:methods}).  The optimal parameter choices are shown in the centre panel.}
\end{figure*}

The creation of the filament mask (Section \ref{section:methods}) is dependent on two parameters: the size of the patch used for the adaptive threshold, and normalization value ($I_0$) used in the arctan transform. The choice of the parameters does not greatly affect the initial detected regions. However, the connectivity between these structure may change drastically. We demonstrate this in Figure \ref{fig:patch_vs_thresh}, where the filamentary structure in the image from Figure \ref{fig:steps} is shown for different combinations of parameters.  Choosing a low normalization value causes the algorithm to ignore faint connections. Choosing a small patch size leads to fragmented regions, which do not adequately describe the large-scale structure. Conversely, a large patch size will lead to ignoring substructure within filamentary structures. The patch size influences the final mask more than the choice of normalization, in fact the mask drastically changes when the patch size is altered by a factor of 2. Optimal results are obtained by setting these parameters to the values discussed in Section \ref{section:methods}. As is shown in Figure \ref{fig:patch_vs_thresh}, bright filamentary structure is largely insensitive to such changes.

The normalization used in the arctan transform is also sensitive to the noise level. When the image is flattened beyond the level of the faint filamentary structure, adaptive thresholding will return large spurious regions devoid of signal. We note that such regions can be identified using the optimal parameters, however they are eliminated by the removal of small objects (see Figure \ref{fig:steps}d-e). Features near the noise level rely on optimal parameter settings to be identified. The sensitivities discussed above are most noticeable in this case and proper identification relies on minimizing fragmentation of the region due to noise.

\pagebreak

\section{Resolution Effects}
\label{app:resolution}

\begin{figure}
\includegraphics[width=0.48\textwidth]{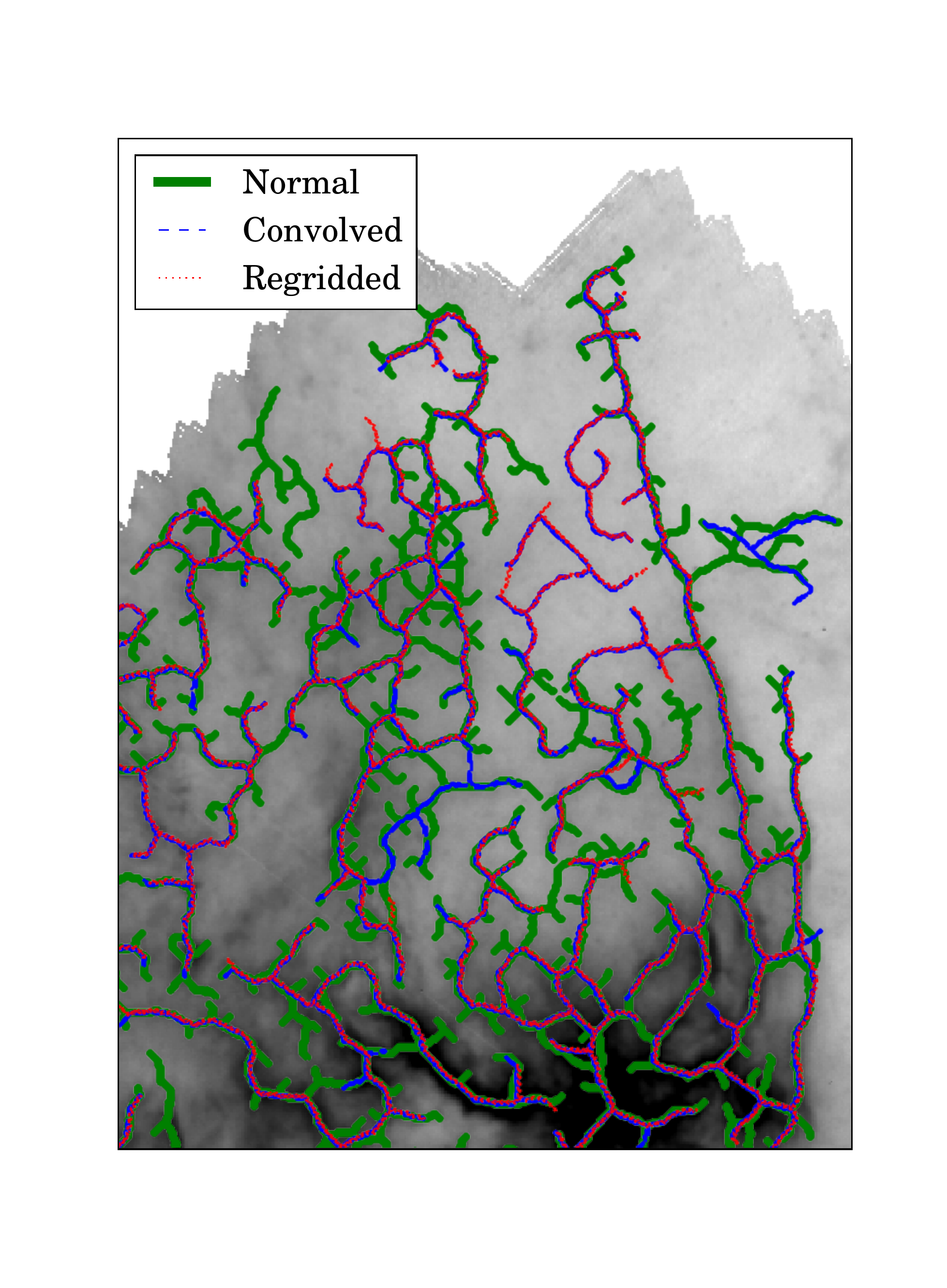}
\caption{\label{fig:pipe_img} Image of the skeleton structures after smoothing the Pipe data to the equivalent physical resolution it would have at 460 pc.  The filamentary structure extracted from the image is nearly the same after degrading the resolution by a factor of 2.5.}
\end{figure}

\begin{figure}
\includegraphics[width=0.48\textwidth]{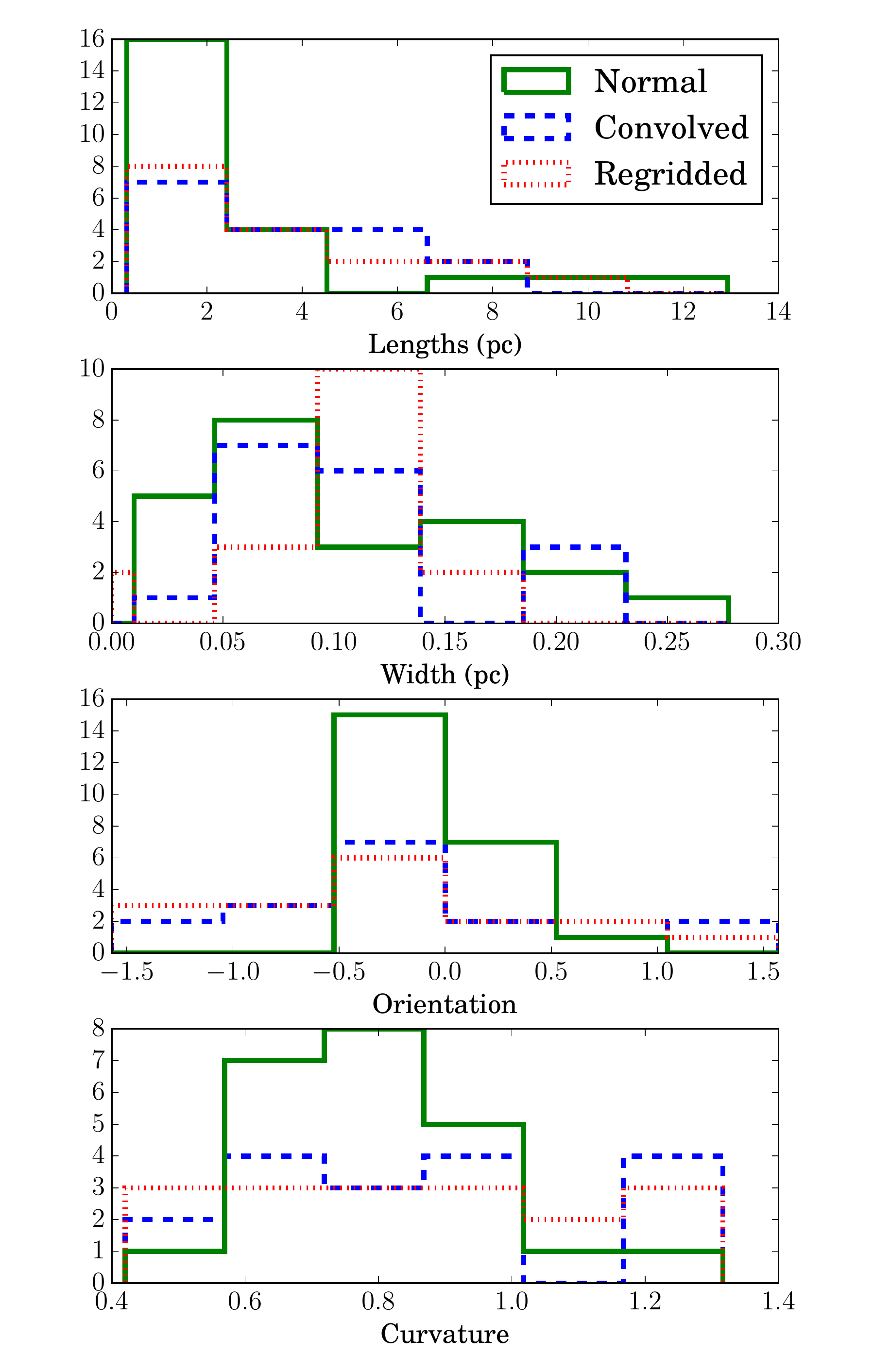}
\caption{\label{fig:pipe_hists} Comparison of the derived skeleton properties shown in \ref{fig:pipe_img}.  The distributions of the properties are similar under this reduction in resolution except that the number of short filaments drops by a factor of two.}
\end{figure}

In this section, we test the derived parameters of the filamentary network with respect to changes in physical resolution.  We complete an analysis of the Pipe region, one of the nearest targets, at its native physical resolution (0.015 pc), at the resolution the data would have were the cloud located at 460 pc (0.056 pc), and the effect of regridding the convolved data back to the original pixel scale.  We process these versions through the {\it FilFinder} algorithm with the same parameters adopted throughout this analysis.  We display the identified filamentary networks in Figure \ref{fig:pipe_img} and the distributions of the properties in Figure \ref{fig:pipe_hists}.

While the network properties and location vary slightly with the degradation of the resolution, they are remarkably stable despite the poorer resolution.  The only noticeable change is that there is a tendency for short filaments to be merged into longer filaments after image degradation.  The width of the filaments do not change significantly once the data are deconvolved.  Overall, the properties do not change significantly within the range of resolutions explored, though this minimal variation is not expected for data sampling dramatically different spatial scales.

Despite this robustness with respect to resolution, we still proceed by degrading all the {\it Herschel} survey data to a common physical resolution to enable a clear comparison between the different systems in the Gould Belt.
\

\label{lastpage}
\end{document}